\documentclass[aps,twocolumn,showpacs,superscriptaddress,pra]{revtex4}

\usepackage{amssymb}
\usepackage{amsmath}
\usepackage{amsfonts}
\usepackage{graphicx}
\usepackage{epsfig}
\usepackage{dcolumn}
\usepackage{bm}
\usepackage{color}
\usepackage{hyperref}
\usepackage[up]{subfigure}
\usepackage{natbib}
\usepackage[normalem]{ulem}

\newcommand{\be}{\begin{equation}}
\newcommand{\ee}{\end{equation}}
\newcommand{\bc}{\begin{center}}
\newcommand{\ec}{\end{center}}
\newcommand{\bea}{\begin{eqnarray}}
\newcommand{\eea}{\end{eqnarray}}

\begin{document}
\title{Effect of control procedures on the evolution of entanglement in open quantum systems}
\author{Sandeep K \surname{Goyal}}
\email{goyal@imsc.res.in}
\affiliation{Optics \& Quantum Information Group, The Institute of Mathematical Sciences, CIT Campus, Taramani, Chennai 600 113, India}

\author{Subhashish \surname{Banerjee}}
\email{subhashish@iitj.ac.in}
\affiliation{Indian Institute of Technology, Rajasthan, Jodhpur 342011, India}

\author{Sibasish \surname{Ghosh}}
\email{sibasish@imsc.res.in}
\affiliation{Optics \& Quantum Information Group, The Institute of Mathematical Sciences, CIT Campus, Taramani, Chennai 600 113, India}

\begin{abstract}
The effect of a number of mechanisms designed to suppress decoherence in open quantum systems are studied with respect to their effectiveness at slowing down the loss of entanglement. 
The effect of photonic band-gap materials and frequency modulation of the system-bath coupling are along expected lines in this regard. However, other control schemes, like resonance 
fluorescence, achieve quite the contrary: increasing the strength of the control kills entanglement off faster. The effect of dynamic decoupling schemes on two qualitatively different system-bath interactions are studied in depth. Dynamic decoupling control has the expected effect of slowing down the decay  of entanglement in a two-qubit system coupled to a harmonic oscillator bath under non-demolition interaction. However, non-trivial phenomena are observed when a Josephson charge qubit, strongly coupled to a random telegraph noise bath, is subject to decoupling pulses. The most striking of these reflects the resonance fluorescence scenario in that an increase in the pulse strength decreases decoherence but also speeds up the sudden death of entanglement. This demonstrates that the behaviour of decoherence and entanglement in time can be qualitatively different in the strong-coupling non-Markovian regime.

\end{abstract}


\maketitle

\section{Introduction}

Entanglement is one of the basic features that distinguish quantum systems from their classical counterparts, and has its origins in the inherent non-locality of quantum mechanics \cite{Bell64}. It is the most useful resource in quantum information theory \cite{Nielsen}, 
and is indispensible for diverse quantum information tasks such as quantum communication, teleportation, quantum error correction, superdense coding, one-way communication etc. In closed systems --- that is, systems which are completely isolated from their surroundings --- entanglement remains conserved under a local unitary evolution,  and can change only under non-local evolution. This makes these systems ideal for quantum information tasks. 
Closed systems are, however, a rarity in the natural world. More often than not, quantum systems are \emph{open}, that is, they are in contact with the surrounding environment --- a thermodynamic reservoir, for example \cite{Louisell73,Caldeira83,Breuer}. Quantum systems are extremely fragile, and the dissipative effects of the environment gives rise to the phenomenon of quantum decoherence \cite{Zurek91, Zurek93}.  As a result, the system undergoes an asymptotic transition to classicality and hence loses all its entanglement, which is a purely quantum phenomenon.  Nevertheless, 
this in itself is not a bad scenario, for if the decoherence rate is low, then entanglement takes a long time to completely disappear and such systems can 
function as useful quantum devices for sufficient periods of time. However, recent studies \cite{YuEbr, Qasimi08} have uncovered systems where the rate of loss of entanglement is exponentially higher than the decoherence rate. This results in a finite time to classicality, and consequently, a finite time to the total loss of entanglement --- a phenomenon given the name entanglement sudden death (ESD). Systems that suffer from ESD are rendered unusable for quantum tasks. Naturally then, ESD has dire implications for the success of quantum tasks, and has become one of the premier branches of quantum information study in recent times. Some of us have recently investigated this phenomenon for the case of $n$-qubit states at finite temperature \cite{SKG10}, as well as for spatially separated $n$-mode Gaussian states coupled to local squeezed thermal baths \cite{SKG10b}.

Given the obvious importance of ESD regarding the 
success of quantum tasks, it is thus a worthwhile exercise to investigate ways and means of controlling the rate of loss of entanglement. Error-correcting codes 
\cite{Calderbank96, Calderbank96b, Shor95, Knill97} and error-avoiding codes \cite{Zanardi97, Duan97, Lidar98} (
which are also known as decoherence-free subspaces) are such attempts. Open loop decoherence control strategies \cite{viola98,Ban98,Duan99,Vitali99, Vitali01,viola99,Agarwal99,Agarwal01,Agarwal01b, Wocjan2002} are another class of widely used strategies used to this effect, where the system of interest is subjected to external, suitably designed, time-dependent drivings that are independent of the system dynamics. The aim is to cause an effective dynamic decoupling of the system from the ambient environment. A comparative analysis of some of these methods has been made in \cite{Facchi04}. An important generalization of the dynamic decoupling scheme, presented recently in \cite{Santos2008}, involves exploiting and merging the randomization and deterministic strategies such as symmetrization, concatenation and cyclic permutation to an $N$ qubit system.  Another mechanism known to slow down the process of decoherence is through manipulation of the density of states. This has been put to use in photonic band-gap materials, which is used to address questions related to the phenomenon of localization of light \cite{John84, Yablonovitch87, John87, Yablonovitch91, John94}. 
 
In this paper, we analyze the evolution of entanglement in two-qubit systems connected to local baths (or reservoirs).  A number of studies of entanglement in  open
quantum systems have been made \cite{Banerjee10,Banerjee10b,Banerjee10c}.  Here we address the need to have a control on the resulting nonunitary evolution, as motivated by the above discussion, and study several methods of doing so. These include manipulation of the density of states in photonic crystals, modulation of the frequency of the system-bath coupling and modulation of external driving on two-qubit systems. A significant part of the paper is devoted to the study of control methods in two-qubit systems undergoing non-Markovian evolution. The first of these is dynamic decoupling --- which is an open-loop strategy --- on a two-qubit system one qubit of which  is in contact with a harmonic oscillator bath. This system undergoes a quantum non-demolition interaction, where dephasing occurs without the system getting damped. The second is a Josephson-junction charge qubit subject to random telegraph ($1/f$) noise due to charge impurities. 

The surprising aspect of this study is that suppression of decoherence due to a control procedure need not necessarily mean preservation of entanglement. In fact, application of resonance fluorescence or dynamical decoupling on the Josephson junction charge qubit, undergoing non-Markovian evolution, results in earlier ESD even though decoherence gets suppressed.

The plan of the paper is as follows. In Section II, we introduce the basic techniques and formalism used in this paper, including the formal way of solving the Lindblad master equation. We also introduce the  concept  of channel-state duality and the factorization law of entanglement decay, both of which will be used subsequently. In Section III, we study the evolution of entanglement in photonic band gap materials and the effect of the special characteristics of such materials on ESD. The effect of frequency modulation of the system-bath coupling on ESD is studied in Section IV. This is followed by a study of  ESD of a two two-level system, one of which is  driven by an external resonant field which is in resonance with the transition frequency. Finally, in section VI, we study the effect of dynamic decoupling on the evolution of entanglement and ESD. We pay particular attention to the Bang-Bang strategy \cite{viola98} with regard to the usual two-qubit system under a QND interaction in section VI (A) and also to a Josephson-junction charge qubit subjected to random telegraph noise, and make comparisons in section VI (B). We conclude our paper in section VII with further discussions. Appendices A, B and C deal with some of the explicit calculations.

\section{Preliminaries}

An open quantum system, as defined in the introduction, is exposed to its environment, which is usually  a thermal reservoir. The dynamics of such a system is naturally dictated by its interaction with its environment. 
If $H$ be the total Hamiltonian of an open system, then $H = H_S + H_R + H_{SR}$, where $H_S$ and $H_R$ are the system and reservoir Hamiltonians respectively and $H_{SR}$ is the interaction Hamiltonian. Open systems undergo nonunitary evolution due to this interaction term, and, depending on the type of the system-reservoir ($SR$) interaction, they can be
broadly divided into two categories --- dissipative and non-dissipative. In the former, the system Hamiltonian does not commute with the interaction Hamiltonian, $[H_S, H_{SR}] \neq 0$, and dephasing occurs along with dissipation and decoherence. In the latter however, these two do commute --- $[H_S, H_{SR}] = 0$ --- and hence the $SR$ interaction is characterized by a class of energy-preserving measurements where dephasing occurs without damping the system \cite{Banerjee07,Banerjee08}. Such a non-dissipative system, as well as the corresponding interaction, is called a Quantum Non-Demolition (QND) system.

We are interested in the time evolution of open quantum system, i.e., of the system state $\rho_S$. Let the initial state of the system-bath combination be $\rho(0)$, and 
let the state at time $t$ be $\rho(t) = U \rho(0) U^{\dagger}$, where $U = e^{-iHt}$ is the time evolution operator. The state of the system alone is obtained from $\rho(t)$ by 
simply tracing out the bath degrees of freedom: $\rho_S(t) = \mathrm{Tr}_R \left[ \rho(t) \right]$, where Tr$_R$ implies a partial trace over the bath. The evolution of the system-bath combination is unitary, and
\begin{align}
\dot{\rho}(t) &= -i[H,\rho(t)]
\end{align}
is the equation of motion. However, the evolution of the system itself is nonunitary, and thus requires a more general equation of motion which, after the application of the Born, Markov and rotating wave approximations, can be written as
\begin{align}
\dot{\rho}_S(t) &= -i[H_S,\rho_S(t)] + \\ \nonumber
& \sum_j\gamma_j\left(F_j\rho_S(t)F_j^{\dagger} -\frac{1}{2}\left\{\rho_S(t),F_j^{\dagger}F_j\right\}\right) .
\end{align}
This is a master equation in its Lindblad form. It can be written in super operator form as
\begin{align}
\dot{\rho}_S(t) &= \mathcal{L}[\rho_S(t)],\\
{\rm and~ in ~matrix ~form,}~~\dot{\rho}_{S\,ij}(t) &= \sum_{k,\,l} L_{ij,kl}\rho_{S\,kl}(t),
\end{align}
where $\mathcal{L}$ is the super operator acting on the system state $\rho_S(t)$ and is effectively a time-derivative, and  $L$ is the
matrix representation of $\mathcal{L}$. In general, $\mathcal{L}$ is time independent and the solution of the above equation can be written
formally as
\begin{align}
\rho_S(t) &= \Lambda[\rho_S(0)]\\
\rho_{S\,ij}(t) &= \sum_{k,\,l} V_{ij,kl}\rho_{S\,kl}(0),
\end{align}
where $V = \exp(Lt)$ is the matrix representation of the time evolution map $\Lambda$. If the system is evolving under unitary
evolution $w$ then the matrix $V$ is simply $w\otimes w^*$, where $w^*$ represents the complex conjugate of $w$ in a fixed basis. 

{\em Channel-State Duality:} A quantum channel is a conduit for the transmission of quantum as well as classical information, and is essentially a completely positive map between spaces of operators. Any such physical quantum channel acting on a $d$-dimensional quantum state can be mapped to a positive operator in $d^2$ dimensions, and, if the channel is trace-preserving, then the corresponding positive operator will have unit trace. Similarly, a valid density matrix in $d^2$-dimensions can be mapped to a trace preserving physical channel acting on  $d$ dimensional systems. Such a two-way mapping between a quantum state and  quantum channel is called channel-state duality. 

The time evolution operator $\Lambda$ is a physical quantum channel represented by the matrix $V$. If $M$ be a valid density matrix corresponding to the map $\Lambda$, it is given by \cite{choi75, jamiolkowski}
\begin{align}
M &= (\mathbb{I}\otimes \Lambda)[|\phi^+\rangle\langle\phi^+|],
\end{align}
where $|\phi^+\rangle = \frac{1}{\sqrt{d}}\sum_{i=1}^d |ii\rangle$ is a maximally entangled state in $d \otimes d$ Hilbert space. Here the channel is applied to one side of the maximally entangled state.

We shall use the symbols $V$ and $M$  for the matrix representation of the time evolution map $\Lambda$ and a valid density matrix corresponding to it, respectively, throughout the paper.

{\em Factorization law of entanglement decay }\cite{Konrad08}: This law says that the evolution of entanglement in a bipartite entangled state under a local one-sided channel can be fully characterized by its action on a maximally entangled state. 
The amount of entanglement at any time $t$ in a given initially entangled two-qubit pure state $|\chi \rangle$, under the action of a one-sided quantum channel, is equal to the product of the initial entanglement in the given state and the entanglement in the state which we get by applying the channel on one side of a two-qubit maximally entangled state. Mathematically this can be written as:
\begin{align} 
\mathcal{C}\left((\mathbb{I}\otimes
\Lambda)[|\chi\rangle\langle\chi|]\right) &=
\mathcal{C}\left(|\chi\rangle\langle\chi|\right) \mathcal{C}\left((\mathbb{I}\otimes
\Lambda)[|\phi^+\rangle\langle\phi^+|]\right),
\end{align}
$\mathcal{C}(.)$ being the concurrence \cite{Hill97,Wootters98}. Therefore, it is enough to study the concurrence in the state obtained after the evolution of $|\phi^+\rangle$ state, i.e, concurrence in the matrix $M$.
We will make use of this factorization law in our subsequent analysis. It can be easily extended to the case where a local quantum channel acts on both the qubits. It should be pointed out that through out the paper, we consider the action of thermal bath as well as the action of the controlling mechanisms on one of the two qubits (say qubit $B$) of a two-qubit system $A+B$. Thus the corresponding action on any two-qubit initial state $\rho_{_{AB}}(0)$ is of the form $(\mathbb{I}\otimes \Lambda_B)(\rho_{_{AB}}(0))$, where $\Lambda_B$ is the associated quantum channel. In case where both the qubits are exposed to individual thermal bath (with or without individual control mechanism for each qubit), the corresponding action will be of the form $(\Lambda_A\otimes \Lambda_B')(\rho_{_{AB}}(0))$, where $\Lambda_A$ is the associated quantum channel acting on qubit $A$ and $\Lambda_B'$ is the associated quantum channel acting on qubit $B$. This is the legitimate quantum operation as the individual qubits are subject to local quantum actions under the associated quantum channels, which does not care whether $\rho_{_{AB}}(0)$ has any entanglement. But one must guarantee that the individual qubits are subject to the action under quantum channels, which is the case in this paper for each of the control mechanisms described.

\section{Evolution of entanglement in the presence of photonic crystals}

In this section we consider a system of two level atoms interacting with a periodic dielectric crystal, this particular structure of which gives rise to the photonic band gap \cite{John84, Ho90, Yablonovitch91}. The effect of this  on electromagnetic waves is analogous to the effect semiconductor crystals have on the propagation of electrons, and leads to interesting phenomena like strong localization of light \cite{John87}, inhibition of spontaneous emission \cite{Yablonovitch91} and atom-photon bound states \cite{John94, Lewenstein88, Yang00}. The origin of such phenomena can ultimately be traced to the photon density of states changing at a rate comparable to the spontaneous emission rates. The photon density of states are of course estimated from the local photon mode density which constitutes the reservoir. It is this photonic band gap that suppresses  decoherence \cite{Wang08}.

 Let us consider a two-qubit system, one qubit of which is locally coupled to a photonic crystal reservoir  kept at zero temperature. In this case,  entanglement dynamics can be obtained by studying  the qubit in contact with the reservoir. We start with the following Hamiltonian:
\begin{align}\label{general-hamil}
H &= \frac{\omega_0}{2} \sigma_z + \sum_k\omega_ka^{\dagger}_ka_k +
\sum_k (g_ka^{\dagger}_k\sigma_- + g_k^*a_k\sigma_+),
\end{align}
where $\omega_0$ is the natural frequency of the two level atom, $\omega_k$ is the energy of the $k$th mode and $g_k$ is the frequency
dependent coupling between the qubit and the photonic crystal, the latter  acting as the reservoir here. And also, $\sigma_2$ and  $\sigma_z$, $\sigma_{\pm} = \sigma_x \pm i\sigma_y$ are the Pauli matrices, with $a_k$ and $a_k^{\dagger}$ being the annihilation and creation operators for the $k$th mode. If we restrict the total atom-reservoir system to the case of a single excitation \cite{garraway97}, the evolution of a given state of the qubit $\rho(0)$ is then given by \cite{Wang08}:
\begin{align}\label{band-gap-master}
\rho(t) &= \left(\begin{array}{cc}
\rho_{11}(0)|c(t)|^2 & \rho_{01}(0)c(t)\\
\rho_{10}(0)c^*(t)& \rho_{00}(0)+\rho_{11}(0)(1-|c(t)|^2)
\end{array}\right),
\end{align}
where
\begin{align*}c(t)=&\varepsilon \left(\lambda_+e^{i\lambda_+^2t}[1 + \Phi(\lambda_+e^{i\pi/4}\sqrt{t})]\right.\\
& \left.- \lambda_-e^{i\lambda_-^2t}[1 + \Phi(\lambda_-e^{i\pi/4}\sqrt{t})]\right),\end{align*}
\[\Phi(x) =\frac{2}{\sqrt{\pi}}\sum_{k=0}^{\infty}\frac{2^kx^{2k+1}}{(2k+1)!!}\mbox{ is the error function },\]
\[\varepsilon = \frac{e^{i\delta t}}{\sqrt{\alpha^2 -4\delta}},\]
\[\lambda_{\pm} = \frac{-\alpha \pm \sqrt{\alpha^2 -4\delta}}{2},\]
\[ \alpha \approx \frac{\omega_0^2d^2}{8\omega_c\epsilon_0(\pi A)^{3/2}}.\]

Here $\delta = \omega_0 - \omega_c$ is the detuning of the atomic frequency and $\omega_c$ is the upper band-edge frequency. 
We have made use of the following photon-dispersion relation near the band edge:~$\omega_k \approx \omega_c + A (k - k_0)^2$, where 
$A \approx \omega_c/k_0^2$, $d$ is the atomic dipole moment and $\epsilon_0$ is the vacuum dielectric constant. 
$\rho (t)$ in Eq. (\ref{band-gap-master}) is the equation of the system taking into account the influence of the reservoir. This
invokes a prescription for the reservoir spectral density, which depends upon the frequency dependent system-reservoir coupling $g_k$, and is codified
in the form of the function $c(t)$, above.

The dynamics $\rho(0)\to\rho(t)$, given by Eq. (\ref{band-gap-master}), is guaranteed to be described by a quantum channel $\Lambda_{pc}$ (say) whose matrix representation is
\begin{align}
V_{pc}& = \left(\begin{array}{cccc}
|c(t)|^2&0&0&0\\
0&c(t)&0&0\\
0&0&c^*(t)&0\\
1-|c(t)|^2&0&0&1\\
\end{array}\right).
\end{align}
Channel-state duality, explained earlier in Section II, ensures that there exists a two-qubit density matrix $M_{pc}$ for every single-qubit channel $V_{pc}$. This matrix $M_{pc}$ can be written as
\begin{align} \label{mband}
M_{pc}& = (\mathbb{I}\otimes\Lambda_{pc})(|\Phi^+\rangle\langle\Phi^+|) \nonumber\\
&=\frac{1}{2}\left(\begin{array}{cccc}
|c(t)|^2&0&0&c(t)\\
0&0&0&0\\
0&0&1-|c(t)|^2&0\\
c^*(t)&0&0&1\\
\end{array}\right),
\end{align}
where $|\Phi^+\rangle = \frac{1}{\sqrt{2}}\left(|00\rangle +|11\rangle\right)$ is a two-qubit maximally entangled state. The concurrence of $M_{pc}$ is $|c(t)|^2$, where  $c(t)$ is a complex-valued function of the detuning parameter $\delta$ and time $t$. If we assume that $\delta = \Delta \alpha^2$, $c(t)$ can then be written in the following simplified form:
\begin{align*}
c(t) = \frac{e^{i\Delta\tau}}{\sqrt{1-4\Delta}}\frac{1}{2}
&\left(d_+e^{id_+^2\tau}[1 
+ \Phi(d_+e^{i\pi/4}\sqrt{\tau})]\right.\\
& \left.- d_-e^{id_-^2\tau}[1 + \Phi(d_-
e^{i\pi/4}\sqrt{\tau})]\right),
\end{align*}
where $d_{\pm} = -1\pm\sqrt{1-4\Delta}$ and $\tau = \alpha^2 t$. Therefore, we need to see the effect of $\alpha$ on entanglement in $M_{pc}$.  Since the entanglement in $M_{pc}$ is $|c(t)|^2$, it is now a function of $\delta$ and $\tau$. 
Invoking the factorization law of entanglement decay, it is sufficient to study entanglement in $M_{pc}$ in order to understand the nature of evolution of entanglement in the two-qubit system.

\begin{figure}
 \subfigure[(Color online) Entanglement (concurrence) of $M$ as a function of $\tau = \alpha^2t$ for
  detuning parameter $\Delta = -0.1$. In this plot the behaviour of
  entanglement is different from the one in the case for $\Delta =
  0$ (see inset). Here (i.e, when $\Delta = -0.1$) entanglement is seen to converge to a non-zero value at
  large $\tau$.]{\includegraphics[width=7.5cm]{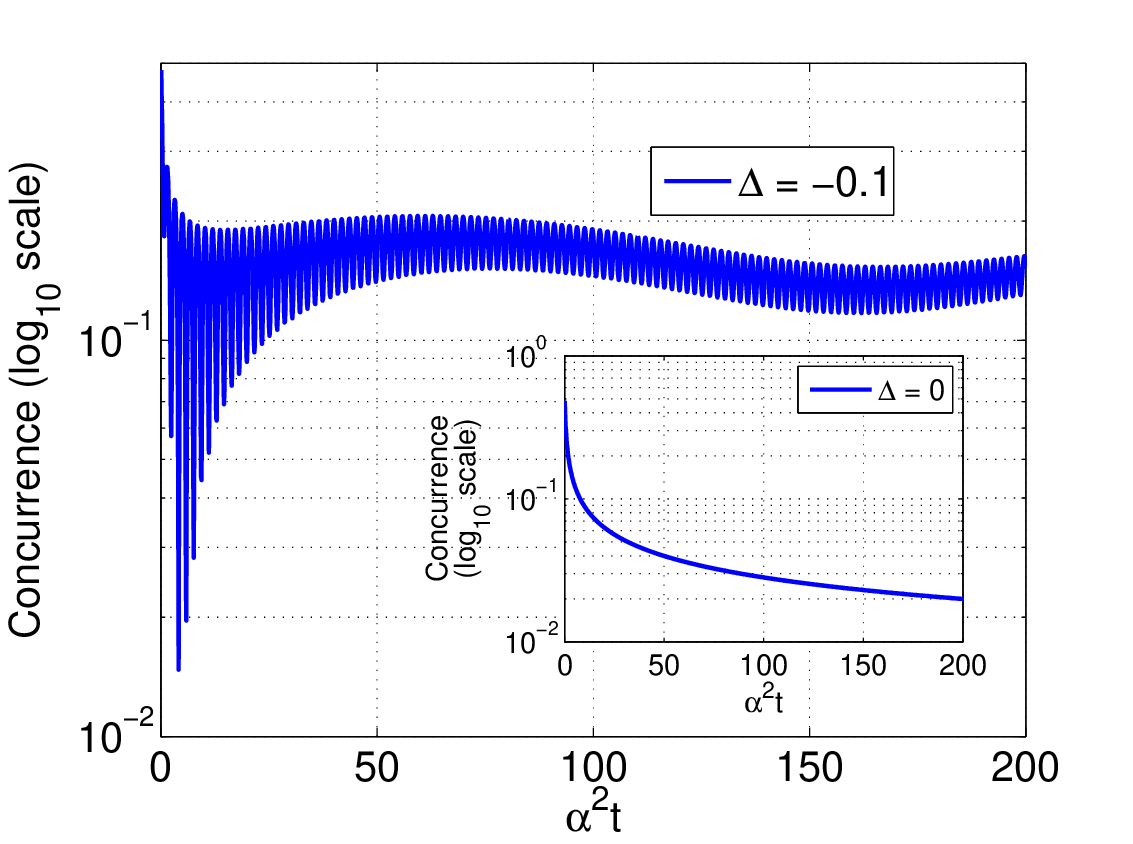}
 \label{band_delta_01}}
 \subfigure[(Color online)Entanglement (concurrence) of $M$ as a function of $\tau = \alpha^2t$ for
  detuning parameter $\Delta = -0.25$. This plot shows that higher the magnitude of
 detuning
 , larger will be the asymptotic value of
 entanglement.]{\includegraphics[width=7.5cm]{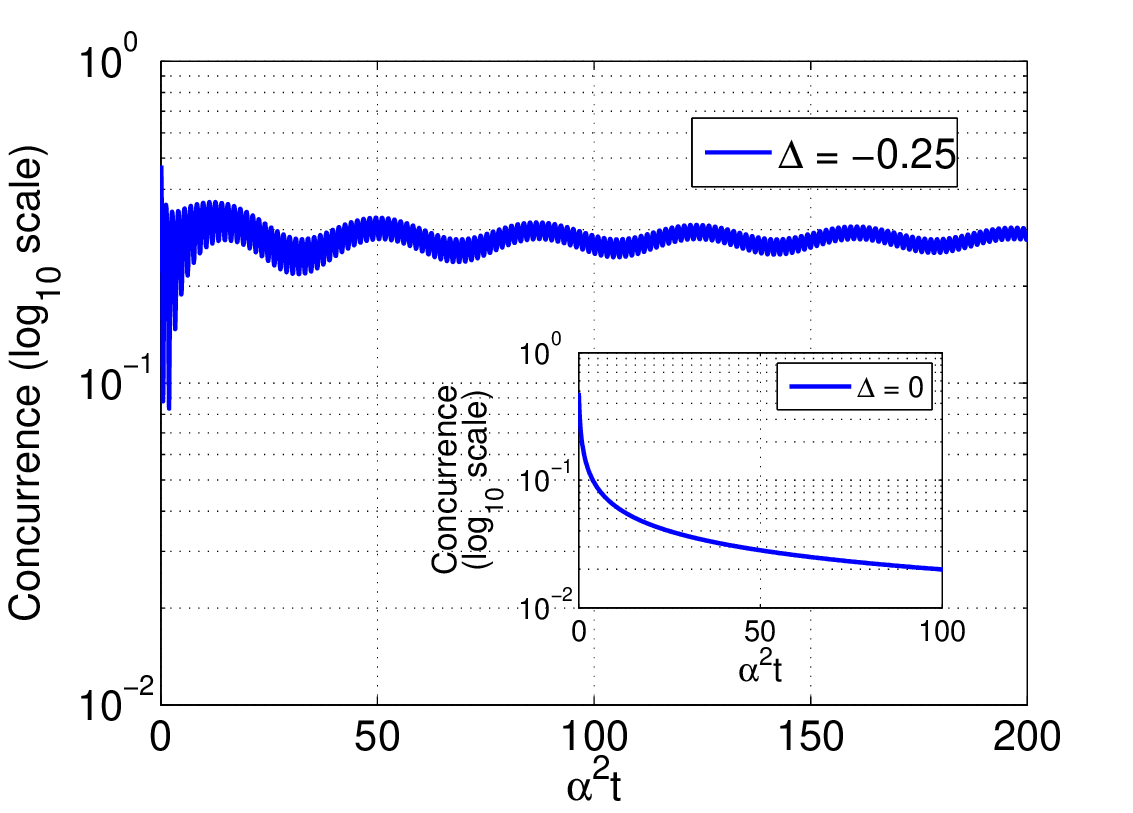} 
 \label{band_delta_025}}
 \caption{Evolution of entanglement in photonic band gap crystals at zero temperature.}\label{bandgap}
 \end{figure}

We show the evolution of entanglement in $M_{pc}$ for different values of $\Delta$ in  FIGS.  \ref{band_delta_01}, \ref{band_delta_025}.  The insets of the figures  depict the evolution of entanglement -- computed using concurrence (see appendix \ref{concurrence}) -- for the usual case of zero band gap, while the main panels show the evolution of entanglement  for increasing influence of the band gap. In FIG.  \ref{band_delta_01}, the system is within a gap in the photonic spectrum, indicated by the negative value of $\Delta$ and hence also $\delta$, as a result of which coherence is preserved and the decay of entanglement is arrested. This feature is further highlighted in FIG.  \ref{band_delta_025}, which is also for the case of negative $\Delta$ of higher order of magnitude than that in FIG.  \ref{band_delta_01}, and as a result there is a greater persistence of entanglement. Thus we find that with the increase of the influence of the photonic band gap on the evolution, entanglement  is preserved longer. From Eq. (\ref{mband}),
it can be seen that, following the arguments of the previous section, there is no ESD in this case, a feature corroborated by the FIGS. \ref{bandgap} provided we choose the initial two-qubit state as a pure entangled state. Since the evolution of the off-diagonal elements of a single-qubit density operator is governed by $c(t)$, the behaviour of the dynamics of coherence is similar to the entanglement dynamics.

Apart from these, Fig. \ref{bandgap} shows another interesting phenomenon -- the temporally damped oscillations in the entanglement. This phenomenon is a signature of the emergence of non-Markovian characteristics in the evolution and implies that the action of detuning changes the character of the dynamics itself, turning it non-Markovian from a Markovian one.

\section{Frequency modulation}

Agarwal and coworkers \cite{Agarwal99,Agarwal01,Agarwal01b} introduced an open-loop control strategy which involved modulation of the system-bath coupling, with the proviso that  the frequency modulation (to be introduced below) should be carried out at a time scale which is faster than the correlation time scale of the heat bath. The technique of frequency modulation has been used earlier to demonstrate the existence of population trapping states in a two-level system \cite{Agarwal94}. Raghavan \emph{et al.} \cite{Raghavan96} showed the connection between trapping in a two-level system under the action of frequency-modulated fields in quantum optics and dynamic localization of charges moving in a crystal under the action of a time-periodic electric field.

Consider the Hamiltonian given in Eq. (\ref{general-hamil}). Frequency modulation essentially involves a modification of the coupling $g_k$ --- the modulated coupling is $g_k\exp\{-i m \sin{\nu t}\}$, where $m$ is the amplitude and $\nu$ is the frequency of the modulation.
The decay of the excited state population can  be significantly arrested by choosing $m$ such that $J_0(m) = 0$, where $J_0$ are the Bessel functions of order zero. The resulting master equation in the interaction picture, when applied to the evolution of a two-level system, is \cite{Agarwal99,Agarwal01,Agarwal01b}: 
\begin{align}
\frac{\partial \rho}{\partial t} =&
-\frac{2(\kappa-i\Delta)J_1^2(m)}{(\kappa-i\Delta)^2+\nu^2}\left\{
C_0^{-+}(\sigma^+\sigma^-\rho - \sigma^-\rho \sigma^+)\right. \nonumber\\
&\left.+ C_0^{+-}(\rho\sigma^-\sigma^+ - \sigma^+\rho
\sigma^-)\right\} + h.c,~~\Delta = (\omega_0 - \omega).\label{masterdriven}
\end{align}
Here $\sigma^{\pm}$ are the Pauli matrices. We have used  the Bessel function expansion $e^{-i m \sin(\nu t)} = \sum_{l= -\infty}^{l = \infty} J_l (m) e^{-i l \nu t}$, where $J_1(m)$ is the Bessel function of order one. Additionally, the modified bath correlation functions are assumed to have the forms $C^{-+} (t) = C^{-+}_0 e^{-\kappa t} e^{i \omega t}$ and $C^{+-} (-t) = C^{-+}_0 e^{-\kappa t} e^{i \omega t}$, where $\kappa$ is the bath correlation frequency. Now, we have 
\begin{align}
\frac{\partial \rho}{\partial t} =& \mathcal{L}_{fm}[\rho], \\
\Rightarrow \rho(t) =& \exp(\mathcal{L}_{fm} t)\rho(0),\\
\Rightarrow \rho(t)_{ij} =& \sum_{kl}\{\exp(\mathcal{L}_{fm}
t)\}_{ij,kl}\rho(0)_{kl} \nonumber\\
=& \sum_{kl}\{V_{fm}(t)\}_{ij,kl}\rho(0)_{kl},
\end{align}
where $V_{fm}(t) = \exp(L_{fm}t)$ and $L_{fm}$ is the matrix representation of $\mathcal{L}_{fm}$. We obtain the matrices $L_{fm}$ and $V_{fm}$ using Eq. (\ref{masterdriven}) and the dynamics turns out to be completely positive. Invoking channel-state duality and the factorization law of entanglement decay, the time to ESD ($t_{ESD}$) is
\begin{align}
t_{ESD} = -\frac{1}{2{\rm Re}(\alpha)T}\log (X_-)
\end{align}
where $\alpha =
\frac{2(\kappa-i\Delta)J_1^2(m)}{(\kappa-i\Delta)^2+\nu^2}$,  $T = C_0^{-+} + C_0^{+-}$ and $X_{-} = \frac{1}{2} \left[ \left( 2+\frac{T^2}{C_0^{-+}C_0^{-+}} \right) - \sqrt{\left(2+\frac{T^2}{C_0^{-+}C_0^{-+}}\right)^2-4} \right]$.
Detailed calculations are given in  Appendix \ref{app-fm}. From FIG. \ref{frequency_modulation}, it can be seen that a higher frequency of modulation sustains entanglement longer.  This result is not altogether surprising, 
for a higher degree of modulation is naturally expected to filter out the influence of the bath and increase the coherence which ultimately results in entanglement sustaining for a longer period of time. And hence the behaviour of the coherence dynamics and entanglement dynamics are qualitatively similar.

 \begin{figure}
 \includegraphics[width=7.5cm]{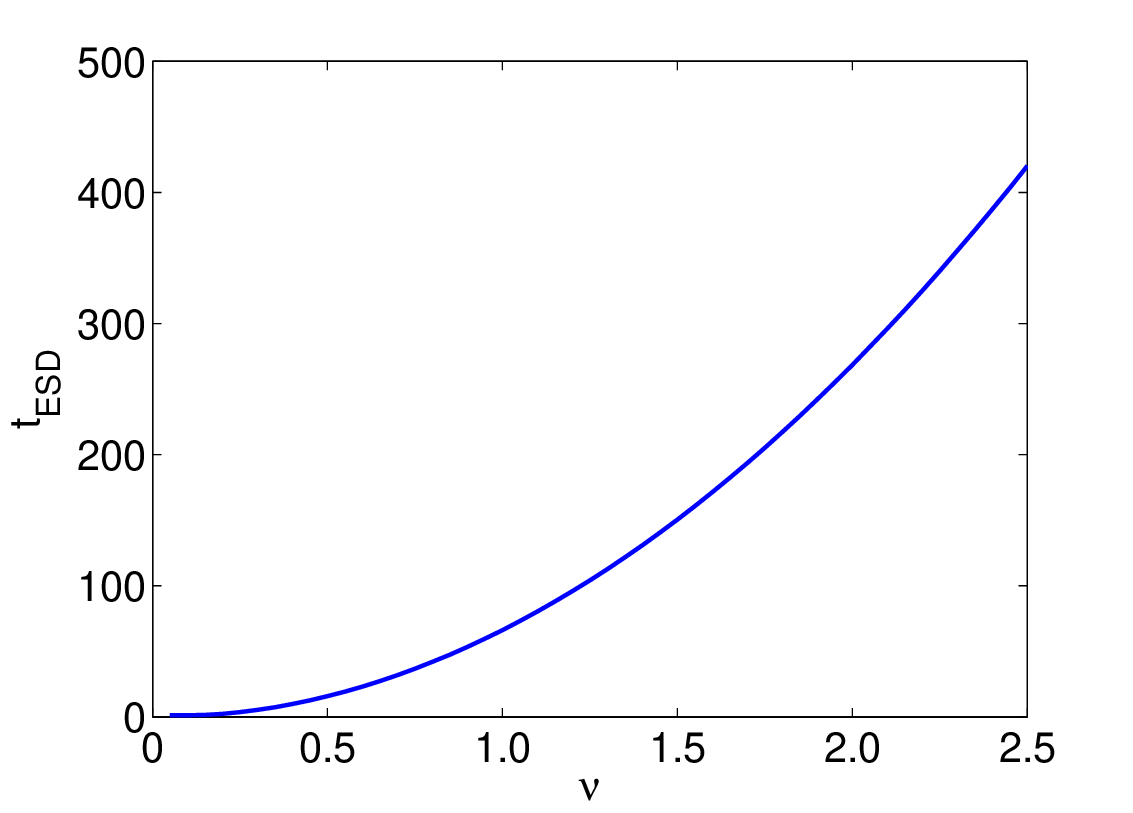}
 \caption{(Color online) In this plot we have time of ESD, i.e,
  the time at which a maximally entangled initial state loses all its
  entanglement and become separable when exposed to a bath, against
  the frequency of modulation $\nu$. In this case we have kept the
  value of $m$ to be the first zero of the Bessel function $J_0$,
  i.e, $m=2.4048$. Also, at $\nu = 0$ the value
  of $t_{ESD}$ is $1.4$. The value of the other parameters are: $\kappa = 0.1$,
  $\Delta = 0.1$, $C_0^{+-} = C_0^{-+} = 0.1$.}\label{frequency_modulation}
 \end{figure}

\section{Resonance fluorescence}\label{resonance-fluorescence}

In the previous section, we focused on the increase in the time to ESD by increasing the degree of frequency modulation of the system-bath coupling. In this section, we study a system where a two-level atomic transition is driven by an external coherent single-mode field which is in resonance with the transition itself. We shall show that, in this situation, an increase in the Rabi frequency --- which plays the role of the modulator --- produces the opposite effect by speeding up ESD. The behavior of such driven systems has been well studied in the literature and has found many applications. In contrast to the situation here, Lam and Savage \cite{Lam94} have investigated a two-level atom driven by polychromatic light. The phenomenon of tunneling in a symmetric double-well potential perturbed by a monochromatic driving force was analyzed by Grossmann \emph{et al.}, \cite{Grossmann91}, while photon-assisted tunneling in a strongly driven double-barrier tunneling diode has been studied by Wagner \cite{Wagner95}.

The analysis of the said driven system begins with its Hamiltonian which, when written in the interaction picture, is $H_{SR} = -E(t) \cdot D(t)$. Here $E(t) = \varepsilon e^{-i\omega_0 t} + \varepsilon^* e^{i \omega_0 t}$ is the electric field strength of the driving mode (treated classically), $\omega_0$ is the atomic transition frequency and $D(t)$ is the dipole moment operator in the interaction picture. The driven two-level system is coupled to a thermal reservoir of radiation modes. If $\gamma_0$ be the spontaneous rate due to coupling with the thermal reservoir and $N = N(\omega_0)$ be the Planck distribution at the atomic transition frequency $\omega_0$, 
 the evolution of this composite system is given by the following master equation \cite{Breuer}:
\begin{align} \label{resonance}
\frac{d}{dt}\rho(t) =& \frac{i\Omega}{2}[ \sigma_+ + \sigma_-,
 \rho(t)]\nonumber\\
&+\frac{\gamma_0(N+1)}{2}\left[ 2\sigma_-\rho(t)\sigma_+
 -\sigma_+\sigma_-\rho(t) - \rho(t)\sigma_+\sigma_-\right]\nonumber\\
&+\frac{\gamma_0(N)}{2}\left[ 2\sigma_+\rho(t)\sigma_-
 -\sigma_-\sigma_+\rho(t) - \rho(t)\sigma_-\sigma_+\right], 
\end{align}
where $\Omega = 2 \varepsilon \cdot d^*$ is the Rabi frequency and $d$ is the transition matrix element of the dipole operator. The term $- \left( \Omega/2 \right) [\sigma_+ + \sigma_{-}]$ characterizes the interaction between the atom and the external driving field in the rotating wave approximation. As usual, $\sigma_{\pm}$ are the atomic raising and lowering operators, respectively.

\begin{figure}
 \includegraphics[width=7.5cm]{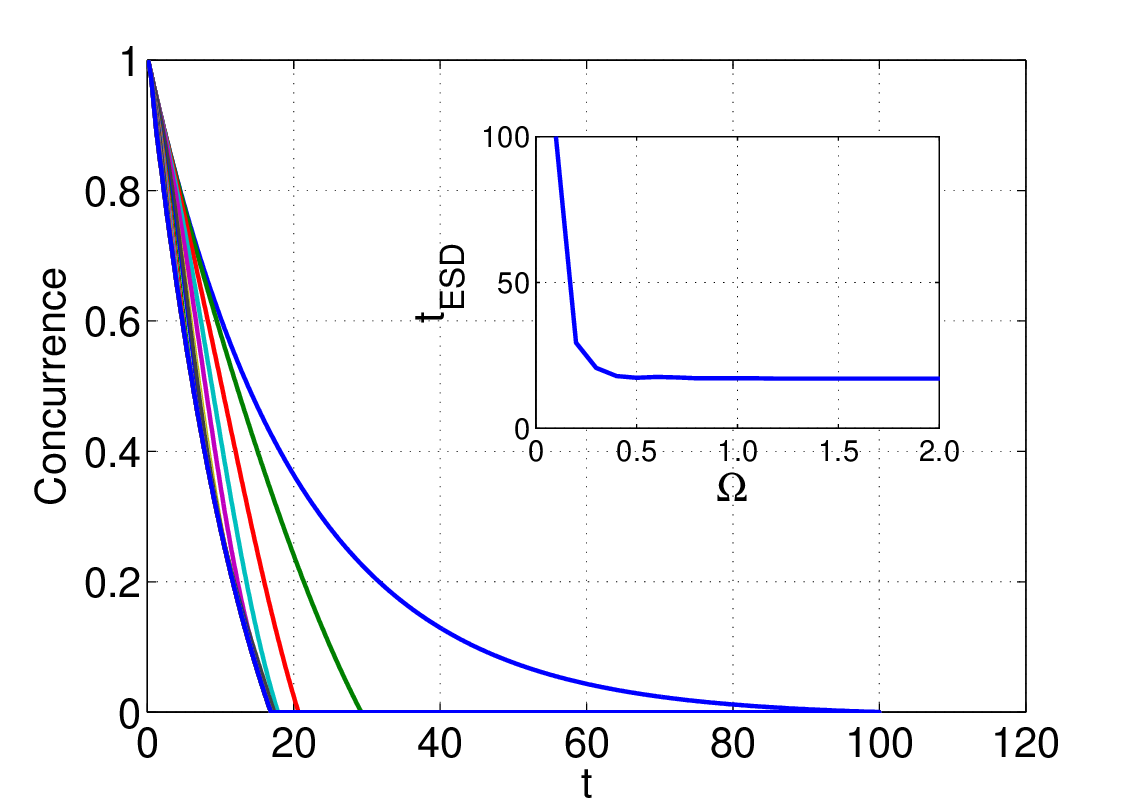}
 \caption{(Color online) Entanglement (concurrence) of $M_{rf}$ as a function of
  time for different values of Rabi frequency $\Omega$, varying from 0
  to 0.5: 0 (the last curve on the right hand side) corresponding 
  to pure damping and 0.5 (first curve on the left hand side)
  corresponding to the underdamped case, i.e., it covers both the 
  overdamped as well as the underdamped cases. In the inset one can
  see that as we increase the $\Omega$ the $t_{ESD}$ seem to converge at $t =
 17.0$. Here $\gamma = 0.1$.} \label{concurrenceVstime} 
 \end{figure}

\begin{figure}
 \includegraphics[width=7.5cm]{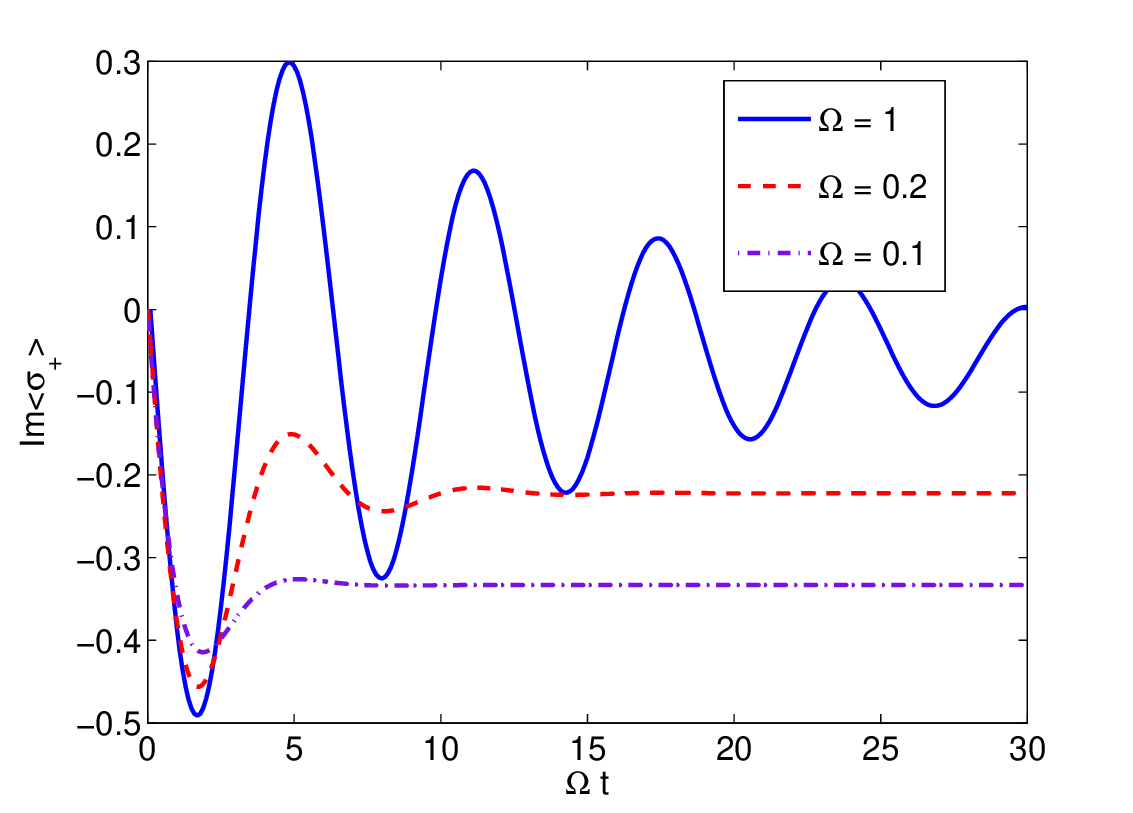}
\caption{(Color online) The plot for the evolution of coherence for a single qubit when the initial state of the qubit is $|\psi\rangle = |0\rangle$ in  the presence of thermal bath. We can see that  the coherence increases as the Rabi frequency $\Omega$ is increased.}\label{coh-rf}
\end{figure}

Let us  consider two identical qubits and, as before, assume that one of them interacts locally with a thermal bath and is subject to monochromatic driving by an external coherent field. The master equation (Eq. \ref{resonance}) yields the corresponding matrices $V_{rf}$ and $M_{rf}$ (where the subscript $rf$ stands for resonance fluorescence) giving rise to a completely positive map:
\begin{align}
V_{rf} &= \left(\begin{array}{cccc}
a_1& a_2& a_2^*& a_4\\
b_1& b_2& b_3& b_4\\
b_1^* & b_3^* & b_2^* & b_4^*\\
d_1& -a_2 & -a_2^* & d_4
\end{array}\right),\\
M_{rf} &= \frac{1}{2}\left(\begin{array}{cccc}
a_1& a_2& b_1& b_2\\
a_2^*& a_4& b_3& b_4\\
b_1^* & b_3^* &d_1& -a_2 \\
b_2^* & b_4^*& -a_2^* & d_4
\end{array}\right),
\end{align}
where
\begin{align*}
a_1 + a_4 =& 1 + \left(1 - X^3\left(\cos(\mu t) -
\frac{\gamma}{4\mu}\sin(\mu t)\right)\right)S_3 \\
&+ \frac{i\Omega}{\mu}X^3\sin(\mu
t)\left(S_- + S_+\right),\\
a_1 - a_4 =& X^3\left[\cos(\mu t) - \frac{\gamma}{4\mu}\sin(\mu
 t)\right],\\
a_2 =& \frac{i\Omega}{\mu}X^3\sin(\mu t),\\
b_1 + b_4 =& -X^2(S_+ + S_-) -\frac{i\Omega}{\mu}X^3\sin(\mu t)S_3\\
&+ X^3\left(\cos(\mu t)
+\frac{\gamma}{4\mu}\sin(\mu
t)\right)(S_--S_+),\\
b_1 - b_4 =& \frac{i\Omega}{\mu}X^3\sin(\mu t),\\
b_{2,3} =& \frac{1}{2}X^2 \pm X^3\left(\cos(\mu t)
+\frac{\gamma}{4\mu}\sin(\mu t)\right),\\
d_1 + d_4 =&2- (a_1+a_4),\\
d_1-d_4 =& -(a_1-a_4),\\
X =& e^{-\frac{\gamma t}{4}},\\
S_+ =& -\frac{i\Omega \gamma_0}{\gamma^2 + 2\Omega^2},\\
S_- =& S_+^*,\\
S_3 =& -\frac{\gamma_0\gamma}{\gamma^2 + 2\Omega^2},\\
\gamma =& \gamma_0(2N + 1),\\
\mu =& \sqrt{\Omega^2 - (\gamma/4)^2}.
\end{align*}
Using these, we plot, in FIG. (\ref{concurrenceVstime}), 
concurrence vs the time to ESD for different values
of the Rabi frequency $\Omega$ and observe that $t_{ESD}$ decreases
for an increase in $\Omega$. This is contrary to the result derived in
the previous section, where an increase in the modulation frequency
$\nu$ delayed the loss of entanglement. The decrease in $t_{ESD}$ does
not however continue indefinitely, but rather saturates to a certain
value for large values of the Rabi frequency. This is an interesting
result, further analysis of which will be carried out in a future
work. Figure (\ref{coh-rf}) depicts an increase in the single-qubit coherence
with an increase in the Rabi frequency $\Omega$, bringing out the fact that here coherence and entanglement behave in a
different fashion.  This puts into perspective the fact that coherence, a local property, need not be monotonic with
entanglement, a non-local property of quantum correlations.

Let us now consider the situation where the system, consisting of the excited two-level atom, is at zero temperature. Let us also consider the evolution of entanglement for two cases demarcated by the relation between the Rabi frequency and the spontaneous rate of coupling with the thermal reservoir. For the underdamped case when $\Omega > \gamma_0/4$, the quantity $\mu$ is real (since $N =0$ at temperature $T = 0$) and hence both the upper level occupation and coherence exhibit exponentially damped oscillations. Conversely, in the overdamped case, $\Omega < \frac{\gamma_0}{4} \Rightarrow \mu $ is purely imaginary and both these quantities decay monotonically to their stationary values. 
The evolution of entanglement, however, works in an opposite way
. Entanglement decays faster for the underdamped case than for overdamping, where the $t_{ESD}$ is higher. 
One possible reason for this could be the relative positions of the three Lorentzian peaks of the inelastic part of the resonance fluorescence spectrum. The central peak is at $\omega = \omega_0$ and the rest are at $\omega = \omega_0 \pm \mu$ \cite{Breuer} for the underdamped case, whereas all three peaks are at $\omega = \omega_0$ for the overdamped case. This indicates that the decay of entanglement in the underdamped should be closely dependent on the quantity $\mu$. This in turn depends on both the dissipation parameter $\gamma$ and the Rabi frequency, the latter in itself a function of the driving strength of the external field and the dipole transition matrix elements.
Thus, in the underdamped case, there exists greater avenues for the decay of quantum coherences as well as entanglement than the overdamped case. Phenomenologically, for the underdamped case ($\Omega > {\gamma}_0/4$), the two-level atom interacts with the external monochromatic field multiple times before spontaneously radiating a photon (see, for example, chapter 10 of ref. \cite{Scully}). Such numerous interactions allows quantum correlations to develop between the two atomic levels and the quantized levels of the field. The phenomenon of monogamy of entanglement \cite{Osborne06} thus ensures that the amount of quantum correlation between the two qubits will decrease. Additionally, it can be seen that at a higher Rabi frequency, $\Omega$ dominates the dissipation and thus causes a saturation of the time to ESD,  as shown in FIG. \ref{concurrenceVstime}.

\section{Dynamic decoupling and the effect on ESD}

As discussed  earlier, open-loop control strategies involve the application of suitably tailored control fields on the system of interest, with the aim of achieving dynamic decoupling of the 
system from the environment \cite{viola98,Vitali99,viola99,Agarwal99,Agarwal01,Agarwal01b}. Bang-Bang control is a particular form of such decoupling where the decoupling interactions are switched on and off at a rate faster than the rate of interaction set by the environment.
The application of suitable radio frequency (RF) pulses, applied fast enough, averages out unwanted effects of the environment and suppresses decoherence.
In this section, we compare the effect of Bang-Bang decoupling on the evolution of entanglement, using channel-state duality and factorization law of 
entanglement decay, in systems connected to two different types of baths. One bath type is composed of infinitely many harmonic oscillators at a finite  temperature $T$ and couples locally to a two-level atom acting as the qubit, while the other adds random telegraph noise to a Josephson-junction charge qubit. It has been shown for the former case that all two-qubit states shows ESD at non-zero $T$ \cite{SKG10}.

\subsection{Bang-Bang decoupling when the bath consists of harmonic oscillators}
\subsubsection{Quantum Non-Demolition Interaction}

Let us consider the interaction of a qubit with a bath of harmonic oscillators where the system Hamiltonian commutes with the interaction Hamiltonian so that there is no exchange of energy between the system and the bath --- this is \emph{quantum non-demolition dynamics} \cite{Banerjee07,Banerjee08}. The only effect of the bath will be on the coherence elements of the qubit evolution, which will
decay in time at the rate $\gamma$. The total Hamiltonian for the system plus bath is:
\begin{align} \label{qnd}
H_0 &= H_q + H_B + H_I;\\
H_q &= \omega_0\sigma_z,\nonumber\\
H_B &= \sum_k \omega_k b^{\dagger}_kb_k,\nonumber\\
H_I &= \sum_k \sigma_z(g_kb^{\dagger}_k + g_k^*b_k).\nonumber
\end{align}
Here the system Hamiltonian $H_q$ commutes with the interaction Hamiltonian $H_I$ and the evolution of such a system is called {\em pure
dephasing}. For simplicity we will work in the interaction picture where the density matrix of the system plus bath and the interaction Hamiltonian transform as:
\begin{align}
\tilde{\rho}(t) &= e^{i(H_q + H_B)t}\rho(t)e^{-i(H_q + H_B)t},\\
\tilde{H}(t) &= \sigma_z\sum_k(g_kb^{\dagger}_ke^{i\omega_k t} + g_k^*b_ke^{-i\omega_kt}).
\end{align}
From here we can write the total time evolution operator for the system plus bath as
\begin{align}
&\tilde{U}(t_0,t) = T\exp\left\{-i\int_{t_0}^tds\tilde{H}(s)\right\} \nonumber\\
&=
\exp\left\{\frac{\sigma_z}{2}\sum_k[b_k^{\dagger}e^{i\omega_kt_0}\xi_k(t-t_0)
  - b_ke^{-i\omega_kt_0}\xi^*_k(t-t_0)\right\},
\end{align}
where $\xi_k(t) = \frac{2g_k}{\omega_k}(1-\exp(i\omega_kt))$. We are interested in calculating 
\begin{align}
\tilde{\rho}_{01}(t) &= \langle  0|{\rm Tr}_B\left\{\tilde{U}(t_0,t)\tilde{\rho}(t_0)\tilde{U}^{\dagger}(t_0,t)\right\}|1\rangle.
\end{align}
Assuming that the bath and the qubit were uncorrelated in the beginning and that the bath is in a thermal state,
we have \cite{viola98}:
\begin{align}
\tilde{\rho}_{01}(t) &= \tilde{\rho}_{01}(t_0)e^{-\gamma(t_0,t)}, 
\end{align}
where 
\begin{align}
\gamma(t_0,t) = \sum_k \frac{|\xi_k(t-t_0)|^2}{2}\coth\left(\frac{\omega_k}{2T}\right).
\end{align}

The matrix representation of the evolution operator $V_{QND}$ (which corresponds to a completely positive map) can be written from here as:
\begin{align}
V_{QND} &= \left(\begin{array}{cccc}
1&0&0&0\\
0&e^{-\gamma(t_0, t)}&0&0\\
0&0&e^{-\gamma(t_0, t)}&0\\
0&0&0&1
\end{array}\right).\label{VQND}
\end{align}
The evolution of maximally entangled state $|\phi^+\rangle = (|00\rangle + |11\rangle)/\sqrt{2}$ provides sufficient information concerning the evolution of entanglement. 
The evolution of one subsystem in state $|\phi^+\rangle$ gives rise to the density matrix:
\begin{align}
M_{QND} &= \frac{1}{2}\left(\begin{array}{cccc}
1&0&0&e^{-\gamma(t_0, t)}\\
0&0&0&0\\
0&0&0&0\\
e^{-\gamma(t_0, t)}&0&0&1
\end{array}\right).
\end{align}
The concurrence in the state $M$ is directly proportional to
$e^{-\gamma(t_0, t)}$.

\subsubsection{Dephasing under Bang-Bang dynamics}

 \begin{figure}
 \includegraphics[width=7.5cm]{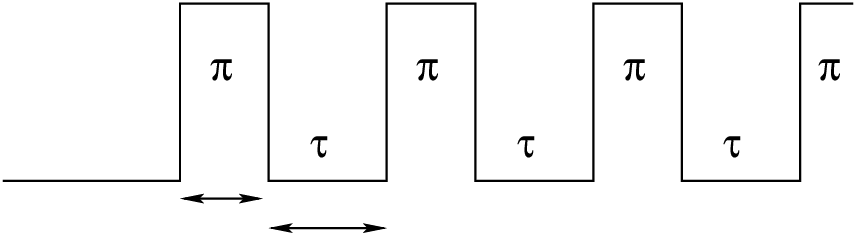}
 \caption{Sketch of the pulse sequence used in bang-bang decoupling
   procedure.} \label{pulse_bang}
 \end{figure}

The function  of Bang-Bang decoupling is to hit  the system of interest with  
a sequence of fast radio-frequency pulses with the aim of slowing down decoherence (see FIG. 
\ref{pulse_bang}). Adding the radio frequency term to the system-plus-bath Hamiltonian $H_0$ (Eq. (\ref{qnd})), we get 
\begin{align}
H(t) &= H_0 + H_{RF}(\omega_0,t ),\\
H_{RF}(t) &=
\sum_{n=1}^{n_p}U^{(n)}(t)\{\cos[\omega_0(t-t_p^{(n)})]\sigma_x + \sin[\omega_0(t-t_p^{(n)})]\sigma_y\}\label{pulse-hamil},
\end{align}
where $t_p^{(n)} = t_0 +n\Delta t,~ n= 1,2,\cdots,n_p$, and
\begin{align}
U^{(n)}(t) &= \left\{
\begin{array}{l}
U~~t_p^{(n)}\le t \le t_p^{(n)} +\tau_p\\
0~~\mbox{elsewhere}.\end{array}\right. \label{bangpulse}
\end{align}
The term $H_{RF}$ acts only on the system of interest which  here is the qubit. It represents a sequence of $n_p$ identical pulses, each of duration $\tau_p$, applied at instants $t = t_p^{(n)}$. The separation between the pulses is $\tau = \Delta t$. The decay rate for this pulsed sequence evolution is \cite{viola98}:
\begin{align}
\gamma_p(N, \Delta t) &= \sum_k\frac{|\eta_k(N,\omega_k\Delta
  t)|^2}{2}\coth\left(\frac{\omega_k }{2T}\right),
\end{align}
where 
\begin{align}
|\eta_k(N, \omega_k \Delta t)|^2 &= 4(1-\cos(\omega_k \Delta t))^2\nonumber\\
&\times\left(N+\sum_{n=0}^{N-1}2n\cos[2(N-n)\omega_k \Delta t]\right).
\end{align}
In \cite{viola98} it has also been shown that $|\eta_k|^2 \le |\xi_k|^2$ which implies that decoherence is suppressed. Also, it is evident that a lower value of $\eta$ implies a lower 
value of $\gamma$. Consequently, we can conclude 
that Bang-Bang decoupling (which corresponds to a new completely positive map, i.e, a modification of the map corresponding to Eq.~\eqref{VQND} due to the RF pulses) slows down entanglement decay.

\subsection{Josephson Junction qubit}

\begin{figure}
\begin{center}
\includegraphics[height=6cm]{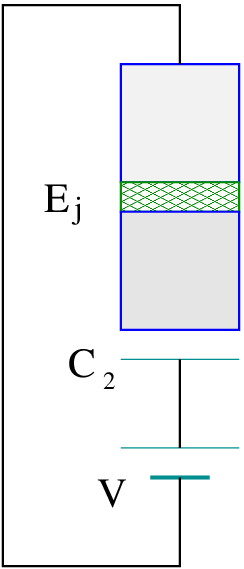}
\caption{(Color online) Schematic diagram for Josephson-junction charge qubit}\label{jj_qubit}
\end{center}
\end{figure}

Although solid state nanodevices satisfy the requirements of large scale integrability and flexibility in design, they are subject to various kinds of low-energy excitations in the environment and suffer from decoherence problems. There have been a number of proposals in this context about the implementation of quantum computers using superconducting nanocircuits \cite{Makhlin99,Falci00}. Experiments highlighting the quantum properties of such devices have already been performed \cite{Nakamura99, Friedman00}. Here the concept of a Josephson-junction qubit comes into prominence. A charge-Josephson qubit is a superconducting island connected to a circuit via a Josephson junction and a capacitor. The computational states are associated with charge $Q$ in the island and are mixed by Josephson tunneling. For temperatures much lower than the Josephson energy, $k_B T \ll E_j$  \cite{Shnirman02,Makhlin01, Paladino02,Paladino03}, we have the Hamiltonian
\begin{align}
H_Q & = \frac{\epsilon}{2}\sigma_z - \frac{E_j}{2}\sigma_x, \label{Josephson}
\end{align}
with the charging energy $E_C$ dominating the Josephson energy. Here, $\epsilon \equiv \epsilon(V) = 4 E_C (1 - C_2 V/e)$, $C_2$ is the capacitance of the capacitor connected to the island and $V$ is the external gate voltage (see Fig. \ref{jj_qubit}). 

Fluctuating background charges (BCs) (charge impurities) are an important source of decoherence in the operation of Josephson charge qubits. 
These are believed to originate in random traps for single electrons in dielectric materials surrounding the superconducting island. At low frequencies, these fluctuations cause the $1/f$ noise 
which is also  known as random telegraph noise, and is directly observed in single electron tunneling devices \cite{Zorin96, Nakamura02}. This has also been studied in the context of fractional 
statistics in the Quantum Hall Effect \cite{Kane03}. This noise, arising out of decoherence, is modeled \cite{Paladino02,Paladino03} by considering each of the BCs as a localized impurity level 
connected to a fermionic band, i.e., the quantum impurity is described by the Fano Anderson model. This is the quantum analogue of the classical model of $N$ independent, randomly activated 
bistable processes. For a single impurity, the total Hamiltonian is:
\begin{align}\label{total-hamiltonian}
H &= H_Q - \frac{v}{2}b^{\dagger}b\sigma_z + H^I,
\end{align}
where
\begin{align}
H^I &= \epsilon_cb^{\dagger}b + \sum_k\left[T_kc_k^{\dagger}b +
  h.c.\right] + \sum_k \epsilon_kc_k^{\dagger}c_k.
\end{align}
Here $H^I$ describes the BC Hamiltonian,  $b$ represents the impurity charge in the localized level $\epsilon_c$, $c_k$  the electron in the band with
energy $\epsilon_k$, and $H_Q$ is as in Eq. (\ref{Josephson}). The impurity electron may tunnel to the band with amplitude $T_k$. 
The BC produces an extra bias $v$ for the qubit via the coupling term $(v/2) b^{\dagger}b\sigma_z$. An important scale is the switching rate $\gamma = 2\pi \rho(\epsilon_c)|T|^2$, 
where $\rho(\epsilon_c)$ is the density of states of the band. It is assumed that we are working in the the relaxation regime of the BC where the tunneling rate to all fermionic bands 
are approximately same, hence $T_k$ gets replaced by $T$, above. The fraction $v/\gamma$ determines whether the operational regime of the qubit is weak ($v/\gamma \ll 1$) or strong ($v/\gamma > 1$). Studying the single BC case is 
important, since it has been shown \cite{Paladino02} that the effect of multiple BCs can be trivially extended from that of a single BC. For multiple strongly coupled BCs producing $1/f$ noise, 
the effect of a large number of slow fluctuators is minimal and pronounced features of discrete dynamics such as saturation and transient behavior are seen. There are two special operational 
points for the qubit related to Eq. (\ref{Josephson}): (a)  $\epsilon = 0$, corresponding to charge degeneracy and (b) $E_j = 0$, for the case of pure dephasing \cite{Bergli06,Abel08}, where 
tunneling can be neglected. We will consider this case  later in detail and make a comparison of ESD, for the case of pure dephasing, between the harmonic oscillator and $1/f$ baths.

The general procedure for  studying the effect of the BC on the dynamics of the qubit is to calculate the unitary evolution of the entire system plus bath and then trace out the bath degree of freedom. Thus, $\rho_Q(t) = {\rm tr_E}\{W(t)\}$, $W(t)$ being the the full density matrix. In the weak coupling limit a master equation for $\rho_Q(t)$ can be written \cite{cohen93}. The results in the standard weak coupling approach are obtained at lowest order in the coupling $v$, but it has been pointed out that higher orders are important for a $1/f$ noise \cite{Shnirman02,Makhlin01,Paladino02}.

The failure of the standard weak coupling approach is due to the fact that the $1/f$ environment includes fluctuators which are very slow on
the time scale of the reduced dynamics. To circumvent this problem one considers another approach in which a part of the bath is treated on
the same footing as the system \cite{Paladino03}. We study the evolution of this new system and later trace out the extra part which belongs to the bath, i.e., 
$\rho (t) = {\rm Tr_{fb}}\{W(t)\}$. We then obtain $\rho_{Q} (t)$ from $\rho (t)$ as $\rho_Q (t) = {\rm Tr_b}\{\rho (t)\}$, where the subscript $fb$ stands for fermionic band. 
In that context we split the Hamiltonian (\ref{total-hamiltonian}) into a system
Hamiltonian $\displaystyle H_0 = H_Q -\frac{v}{2} \, b^{\dagger} b\sigma_z + \epsilon_cb^{\dagger}b $ and environment Hamiltonian $H_E = \sum_k\epsilon_kc^{\dagger}_kc_k$ 
coupled by $\mathcal{V} = \sum_k\left[T_kc^{\dagger}_kb + h.c.\right]$. The eigenstates of $H_0$
are product states of the form $|\theta\rangle|n\rangle$, e.g,
\begin{align*}
|a\rangle &= |\theta_+\rangle|0\rangle,\\
|b\rangle &= |\theta_-\rangle|0\rangle,\\
|c\rangle &= |\theta'_+\rangle|1\rangle,\\
|d\rangle &= |\theta'_-\rangle|1\rangle,
\end{align*}
with corresponding energies
\begin{align*}
-\frac{\Omega}{2},\frac{\Omega}{2},~ -\frac{\Omega'}{2}+\epsilon_c,\frac{\Omega'}{2}+\epsilon_c.
\end{align*}
Here $|\theta_{\pm}\rangle$,$|\theta_{\pm}'\rangle$ are the two eigenstates of $\sigma_{\hat{n}}$,~$\sigma_{\hat{n}'}$ respectively, the directions $\hat{n},\hat{n}'$ being specified by the polar angles $\theta$ with $\phi = 0$ and $\theta'$ with $\phi'=0$. The two level  splittings are $\Omega = \sqrt{\epsilon^2 + E_j^2}$ and $\Omega' = \sqrt{(\epsilon+v)^2 + E_j^2}$, and $\cos(\theta)= \epsilon/\Omega,~\sin(\theta)=E_j/\Omega,~\cos(\theta') = (\epsilon+v)/\Omega',~ \sin(\theta') = E_j/\Omega'$. Here $b^{\dagger}b|0\rangle =0$ and  $b^{\dagger}b|1\rangle =|1\rangle$.

The master equation for the reduced density matrix $\rho (t)$ in the Schr\"{o}dinger representation and in the basis of the eigenstates of $H_0$ reads:
\begin{align}
\frac{d\rho_{ij}(t)}{dt} &= -i\omega_{ij}\rho_{ij}(t) + \sum_{mn}R_{ij,mn}\rho_{mn}(t), \label{masterPaladino}
\end{align}
where $\omega_{ij}$ is the difference of the energies ($\omega_{ab} = \Omega/2 - (-\Omega/2)$, etc.) and $R_{ij,mn}$ are the elements of the Redfield tensor \cite{cohen93} where $i,j = \{ a,b,c,d\}$. These are given by
\begin{align}
R_{ij,mn} &= \int_0^{\infty}d\tau \Bigg\{ c^>_{njmi}(\tau)e^{i\omega_{mi}\tau}+c^<_{njmi}(\tau)e^{i\omega_{jn}\tau}\nonumber\\
&\left.-\delta_{nj}\sum_kc^>_{ikmk}(\tau)e^{i\omega_{mk}\tau}-\delta_{im}\sum_kc^<_{nkjk}(\tau)e^{i\omega_{kn}\tau}\right\},
\end{align}
where
\begin{align}
c_{ijkl}^{\gtrless}(t) &= [\langle i |b|j\rangle\langle
  l|b^{\dagger}|k\rangle + \langle i |b^{\dagger}|j\rangle\langle
  l|b|k\rangle]iG^{\gtrless}(t).
\end{align}

Here $iG^>(\omega)=\gamma/(1-e^{-\beta\omega})$ is the Fourier transform of $G^>(t)$ and $G^<(\omega) = G^>(-\omega)$, therefore, $G^<(t) = G^>(-t)$. This problem has a very interesting symmetry: 
the diagonal and off diagonal elements do not mix if the initial state $\rho(0)$ is a diagonal density matrix in the BC. Therefore, we can divide the Redfield tensor elements in two parts, one  corresponding to population (diagonal elements) and other  corresponding to 
coherence (off diagonal elements).

The $R_{ii,nn}$  elements which affect the population are:
\begin{align}
R_{ii,nn} &= \int_0^{\infty}\left\{ \chi_{in}iG^>(\tau)e^{i\omega_{ni}\tau}
+ \chi_{in}iG^<(\tau)e^{-i\omega_{ni}\tau}\right\}\nonumber\\
&= \chi_{in}\left[ iG^>(\omega_{ni}))\right].
\end{align}
Here  $n \ne i$ and $\chi_{in} = (|\langle n|b|i\rangle|^2 + |\langle
n|b^{\dagger}|i\rangle|^2)$, and 
\begin{align}
R_{ii,ii}&= -\sum_k\chi_{ik}\left[ iG^>(\omega_{ik})\right].
\end{align}

Now we  calculate the elements which are responsible for the coherence part. In the adiabatic regime we have $\gamma \sim \Omega - \Omega' \ll \Omega$ and $ \Omega'$, i.e., where the BCs are not static and the mixing of $\rho_{ab}$ and  $\rho_{cd}$ in Eq.~(\ref{masterPaladino}), as well as their conjugates cannot be neglected. Hence the non-zero elements of $R$ tensor -- which affect  coherence -- are the following:
\begin{align*} 
R_{ab,ab} &= -\frac{\gamma}{2}\left[1-c^2\delta -s^2\delta' +i(c^2w + s^2w')\right],\\
R_{cd,cd} &= -\frac{\gamma}{2}\left[1+c^2\delta +s^2\delta' +i(c^2w - s^2w')\right],\\
R_{ab,cd} &= \frac{c^2\gamma}{2}\left[1+\delta -iw\right],\\
R_{cd,ab} &= \frac{c^2\gamma}{2}\left[1-\delta -iw\right].
\end{align*}
Here 
\begin{align*}
c &= \cos[(\theta-\theta')/2],\\
s &= \sin[(\theta-\theta')/2],\\
\delta &= t_{ca} + t_{db},\\
\delta'&= t_{da} + t_{cb},\\
w &= w_{ca}-w_{cb},\\
w' &= w_{da} - w_{cb},\\
t_{ij} &=\frac{1}{2}\tanh\left(\frac{\beta\omega_{ij}}{2}\right),\\
w_{ij}&= -\frac{1}{\pi}{\rm Re}\left\{\psi\left(\frac{\pi +i\beta\omega_{ij}}{2\pi}\right)\right\} ,
\end{align*}
and $\psi(z)$ is the digamma function. 

Now we can construct the explicit form of the matrix $R = [R_{ij,mn}]$ 
\begin{widetext}
\begin{align}
R &= \left(\begin{array}{cccc|cccc|cccc|cccc}
R_{1,1}&0&0&0&0&R_{1,2}&0&0&0&0&R_{1,3}&0&0&0&0&R_{1,4}\\
0&z_-&0&0&0&0&0&0&0&0&0&y_+&0&0&0&0\\
0&0&0&0&0&0&0&0&0&0&0&0&0&0&0&0\\
0&0&0&0&0&0&0&0&0&0&0&0&0&0&0&0\\
\hline
0&0&0&0&z_-^*&0&0&0&0&0&0&0&0&0&y_+^*&0\\
R_{2,1}&0&0&0&0&R_{2,2}&0&0&0&0&R_{2,3}&0&0&0&0&R_{2,4}\\
0&0&0&0&0&0&0&0&0&0&0&0&0&0&0&0\\
0&0&0&0&0&0&0&0&0&0&0&0&0&0&0&0\\
\hline
0&0&0&0&0&0&0&0&0&0&0&0&0&0&0&0\\
0&0&0&0&0&0&0&0&0&0&0&0&0&0&0&0\\
R_{3,1}&0&0&0&0&R_{3,2}&0&0&0&0&R_{3,3}&0&0&0&0&R_{3,4}\\
0&y_-&0&0&0&0&0&0&0&0&0&z_+&0&0&0&0\\
\hline
0&0&0&0&0&0&0&0&0&0&0&0&0&0&0&0\\
0&0&0&0&0&0&0&0&0&0&0&0&0&0&0&0\\
0&0&0&0&y_-^*&0&0&0&0&0&0&0&0&0&z_+^*&0\\
R_{4,1}&0&0&0&0&R_{4,2}&0&0&0&0&R_{4,3}&0&0&0&0&R_{4,4}
\end{array}\right),\label{redfield-tensor}
\end{align}
\end{widetext}

where
\[ z_- = -\frac{\gamma}{2}\left[1-c^2\delta -s^2\delta' +i(c^2w + s^2w')\right],\]
\[z_+ = -\frac{\gamma}{2}\left[1+c^2\delta +s^2\delta' +i(c^2w - s^2w')\right],\]
\begin{align*}
y_+ &= \frac{c^2\gamma}{2}\left[1+\delta -iw\right],\\
y_- &= \frac{c^2\gamma}{2}\left[1-\delta -iw\right],\\
R_{i,j} &= R_{ii,jj}.
\end{align*}
 
 Channel-state duality implies that the exponential of the matrix $R$ is the matrix representation of  the evolution channel after neglecting the term $-i\omega_{ij}\rho_{ij}(t)$ in Eq.~\eqref{masterPaladino}. Therefore, we have $V = \exp(Rt)$ which gives us the evolution for the qubit plus the charge impurity. From here we can find the evolution map $V_s$ acting on the qubit (see Appendix \ref{app-tele} for more details) which corresponds to a completely positive map.

 \begin{figure}
 \includegraphics[width=7.5cm]{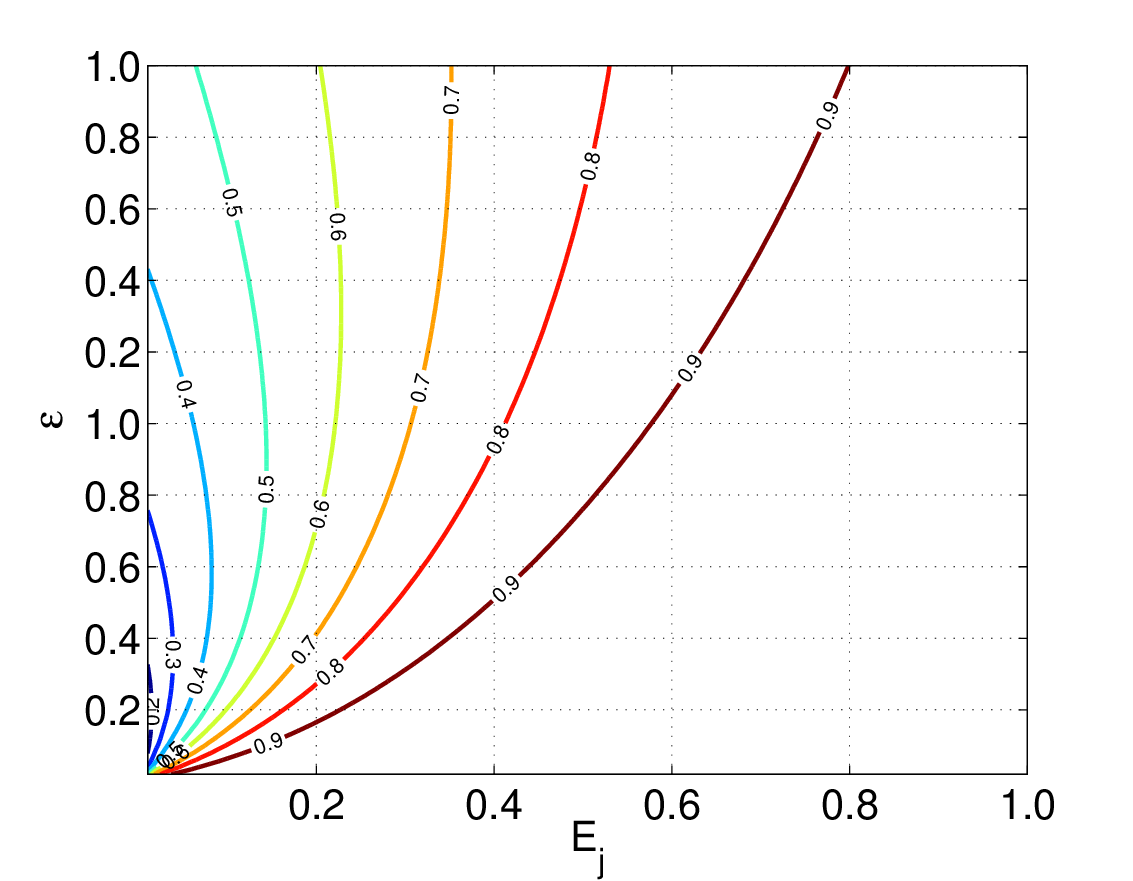}
 \caption{(Color online) Contour of the entanglement after time  $t = 5$ for all
   values of $E_j$ and $\epsilon$ (Josephson junction Hamiltonian parameters). 
   The temperature in this case is equal to $ 0$, $\gamma = 1$,
   $\kappa = v/\gamma = 0.45$.}\label{epsilon-vs-Ej-ent} 
 \end{figure}

\begin{figure}
 \includegraphics[width=7.5cm]{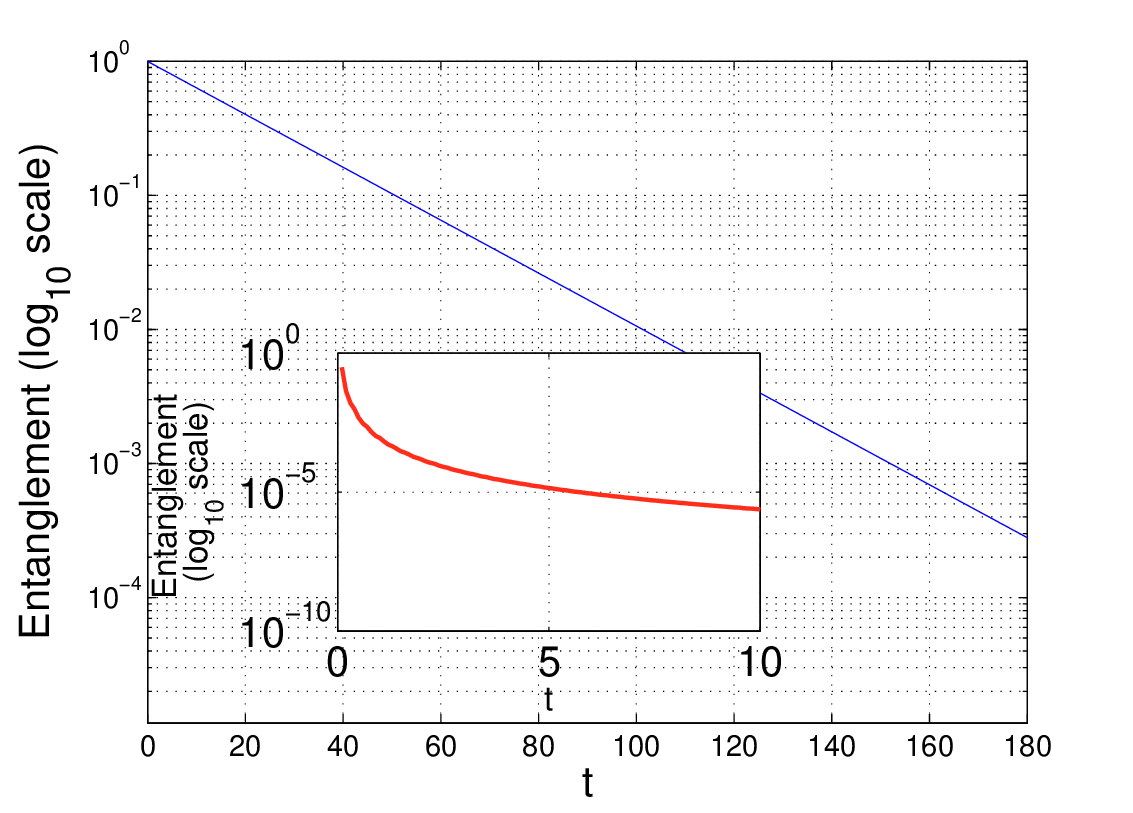}
 \caption{(Color online) Evolution of entanglement  for the pure dephasing case, i.e,
   $E_j = 0$. We can see that whereas  entanglement
   decay is exponential for  the 1/f noise,  it  slows down for a bath  of harmonic oscillators (inset) at zero temperature. }\label{dephasing_telegraph}
 \end{figure}

\begin{figure}
 \includegraphics[width=7.5cm]{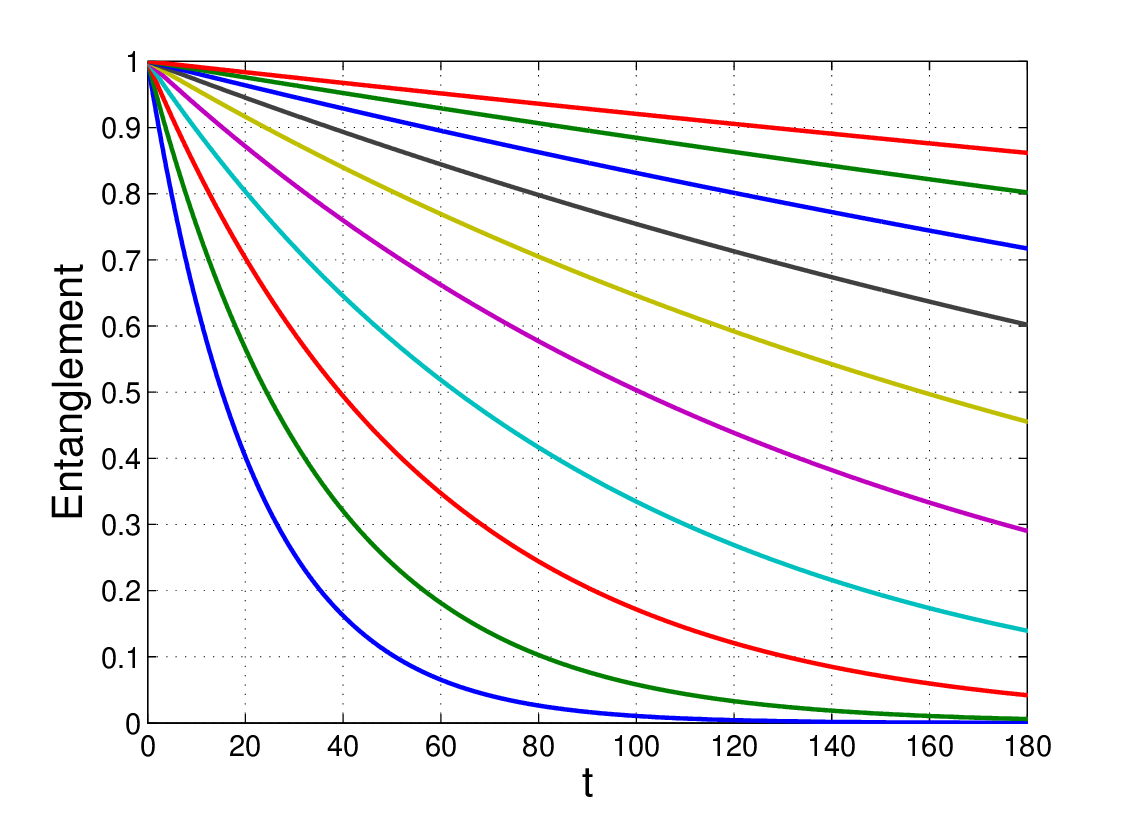}
 \caption{(Color online) Evolution of entanglement
    for $1/f$ (telegraph) noise at zero temperature. Here different curves
   represent the evolution of  entanglement for different values of
   $E_j$ (from $0$ to $1$), with the curve at the bottom corresponding to $E_j = 0$ and that at the top to $E_j = 1$, while $\epsilon$ is fixed and equal to $1$.}\label{telegraph_esd} 
 \end{figure}

The two parameters $\epsilon$ and $E_j$ in the Hamiltonian for the charge Josephson qubit $H_Q$, given in Eq.~\eqref{Josephson}, play a crucial role in the decoherence   properties of the system. For example, if $E_j = 0$, 
the system Hamiltonian $H_Q$ commutes with the interaction Hamiltonian. This situation, as mentioned earlier,  is called non-demolition evolution or pure dephasing. 
In this case there is no energy exchange between system and  bath. On the other hand when we have $\epsilon = 0$, the system Hamiltonian does not commute with the interaction Hamiltonan. Therefore, the two situations are qualitatively different. We present, in FIG.  \ref{epsilon-vs-Ej-ent} a plot of entanglement in the phase space of $\epsilon$ and $E_j$ by evolving a maximally entangled state of two qubits with the bath (of charge impurities) acting only on one qubit. The qubit is evolved for a fixed time $t$ and the entanglement is calculated for different  values of $\epsilon$ and $E_j$. FIG.  \ref{epsilon-vs-Ej-ent} shows that the entanglement in the system increases with an increase in $E_j$ when $\epsilon$ is held fixed, but decreases with an increase in $\epsilon$ when $E_j$ is held fixed.  This is counterintuitive because dissipation increases with the increase in the value of $E_j$. FIG. \ref{dephasing_telegraph} compares the time-evolution of entanglement for the harmonic oscillator bath with the charge-impurities bath, both under pure dephasing.
While entanglement decay is exponential for the case of $1/f$ noise, it is slower for a bath of harmonic oscillators.  
We compare, in FIG.  \ref{telegraph_esd}, the time-evolution of entanglement for various values of Josephson energy ($E_j$) starting with the pure dephasing case given by $E_j = 0$. We see that the entanglement remaining in the system increases with an increase in the value of Josephson energy. This is consistent with FIG. \ref{epsilon-vs-Ej-ent}.


\begin{figure}
 \subfigure[(Color online) Plot of the  entanglement
   (concurrence) as a function of time (t) for different values of coupling strength
   $\kappa=v/\gamma$,  in the weak coupling
 regime, i.e, $\kappa\ll 1$. Here  the range of $\kappa$ is from $0.05$
 to $0.5$, with $0.05$ corresponding to the uppermost curve, and $0.5$ to the lowest (bottom) one. We can see that as we increase $\kappa$, entanglement decreases.]
 {\includegraphics[width=7.5cm]{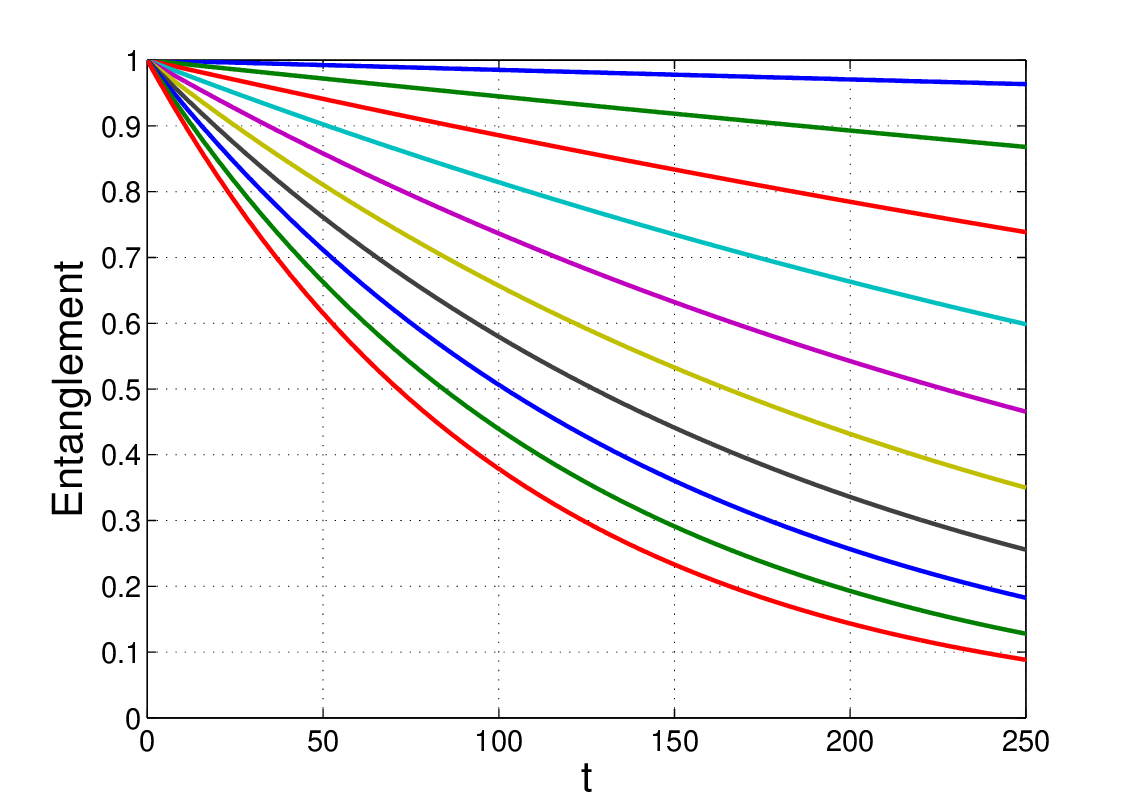}
 \label{kappa_time_weak}}
 \subfigure[(Color online) Plot of the  entanglement
   (concurrence) as a function of time (t) for different values of coupling strength
   $\kappa=v/\gamma$,  in the strong coupling
 regime, i.e, $\kappa > 1$. Here  the range of $\kappa$ is from $5.05$
 to $5.5$. We can see that in the strong coupling region all the curves
 converge.]
 {\includegraphics[width=7.5cm]{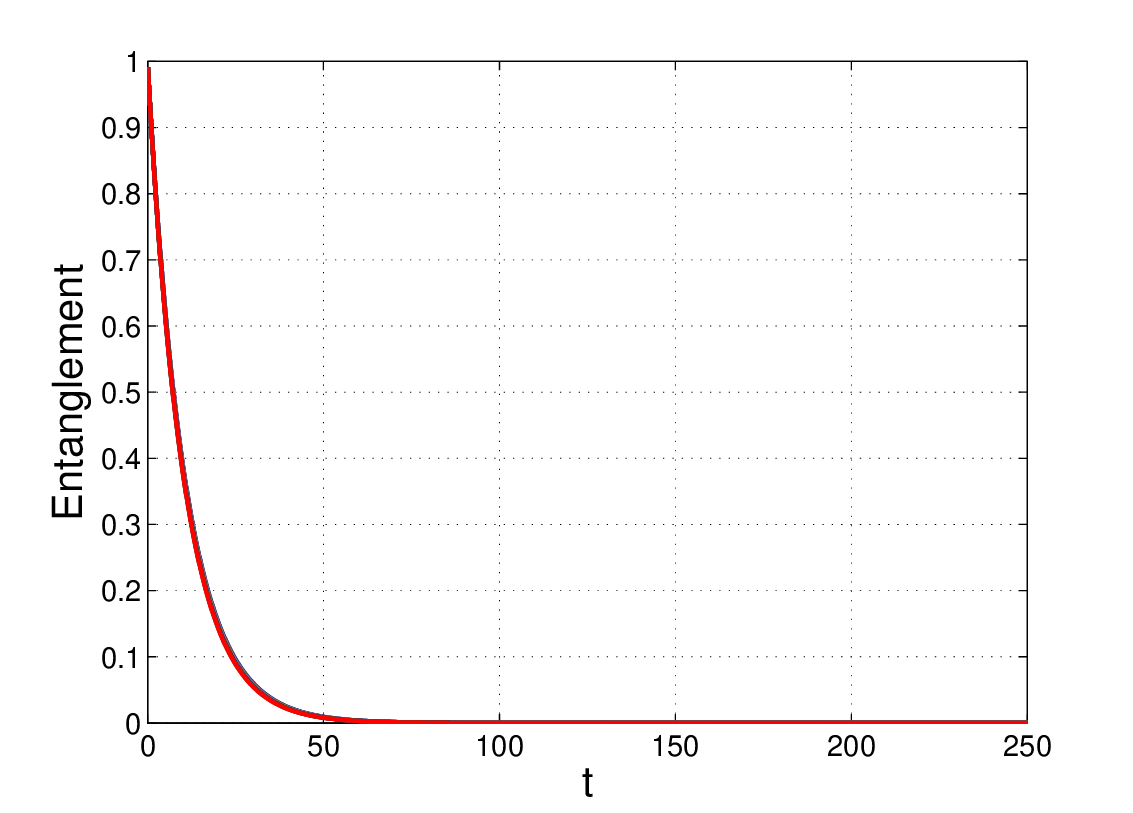}
 \label{kappa_time_strong}}
 \caption{Evolution of entanglement with respect to time, for different coupling strengths and temperature $T = 0$, $E_j = 1 = \epsilon$.}\label{entkappa}
 \end{figure}

Decoherence produced by background charges depends qualitatively on the ratio $\kappa = v/\gamma$, where $\kappa \ll 1$ denotes the weak-coupling regime and $\kappa > 1$ is the strong coupling regime. The latter gives rise to qualitatively new properties.  We  find that (see FIG. \ref{kappa_time_strong}), for $\kappa > 1$,  the time-evolution of entanglement does not depend on $\kappa$. This is in contrast to the weak coupling regime, where the time-evolution of entanglement does depend on $\kappa$, as seen in FIG.  \ref{kappa_time_weak}, where an increase in $\kappa$ leads to a decrease in entanglement. Naturally, decoherence due to the bath forces entanglement to decay with time for both cases.

\subsubsection{Evolution operator with Bang-Bang interaction}

\begin{figure}
 \includegraphics[width=7.5cm]{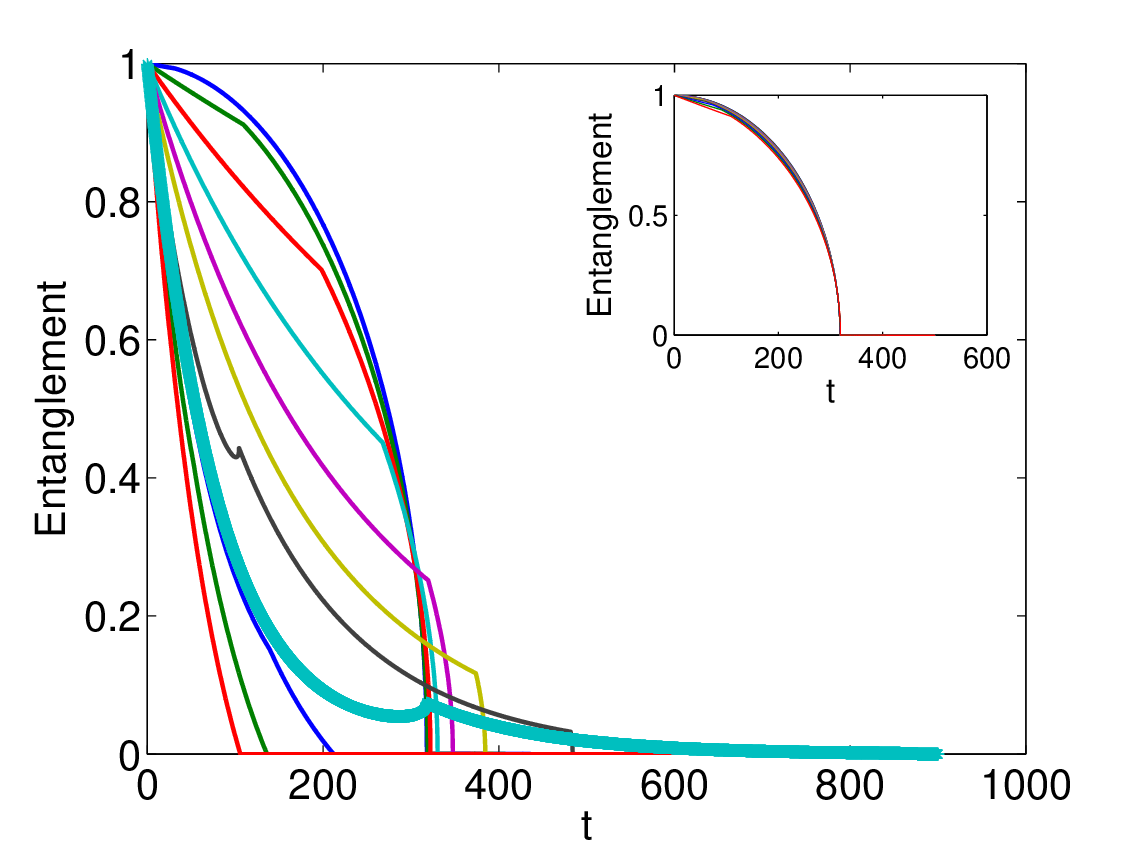}
 \caption{(Color online) Effect of bang-bang decoupling on
 entanglement, in the weak coupling regime. If we compare this plot
 with FIG. (\ref{kappa_time_weak}), with the curves corresponding to the same values of $\kappa$, we  see that the bang-bang
 decoupling causes  entanglement to disappear faster in time for a fixed
 value of the coupling strength. Here the parameters are same as in
 FIG. (\ref{kappa_time_weak}) and the pulse strength is $U = 50\pi$ with
 time for which the pulse was activated is $\tau_p = 0.01$. In the
 inset we have the evolution of entanglement for very small range
 ($0.01$ to $0.1$) of coupling $\kappa$. The thickest curve is the
 one corresponding to $\kappa = 0.38$. This curve is important in the sense
that it has the largest $t_{ESD}$.}\label{bang-weak-kappa} 
 \end{figure}

\begin{figure}
\includegraphics[width=7.5cm]{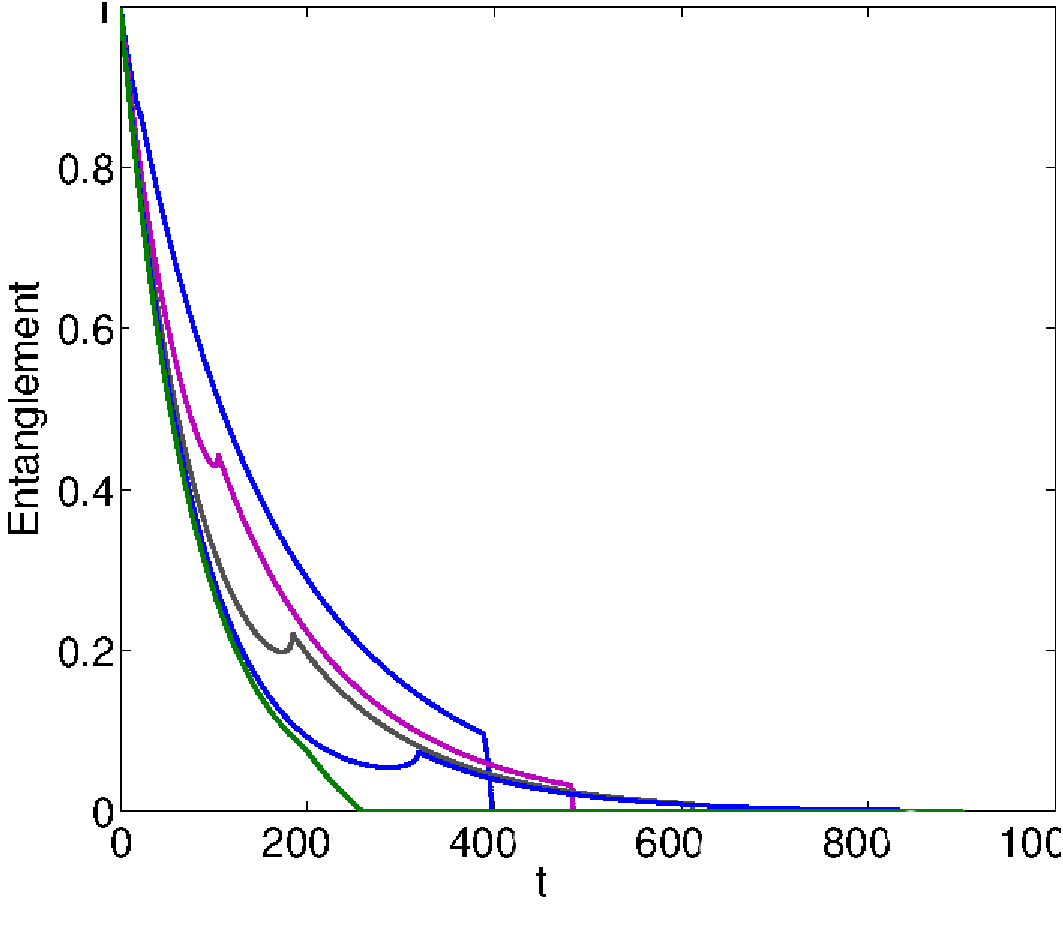}
\caption{(Color online) Entanglement evolution for the coupling parameter range
  $0.3<\kappa<0.4$, with the uppermost curve corresponding to $\kappa = 0.3$ and the lowest (bottom) curve corresponding to 0.4. One can see from this plot the formation and
  disappearance of the kink.}\label{bang-kink}
\end{figure}

\begin{figure}
 \includegraphics[width=7.5cm]{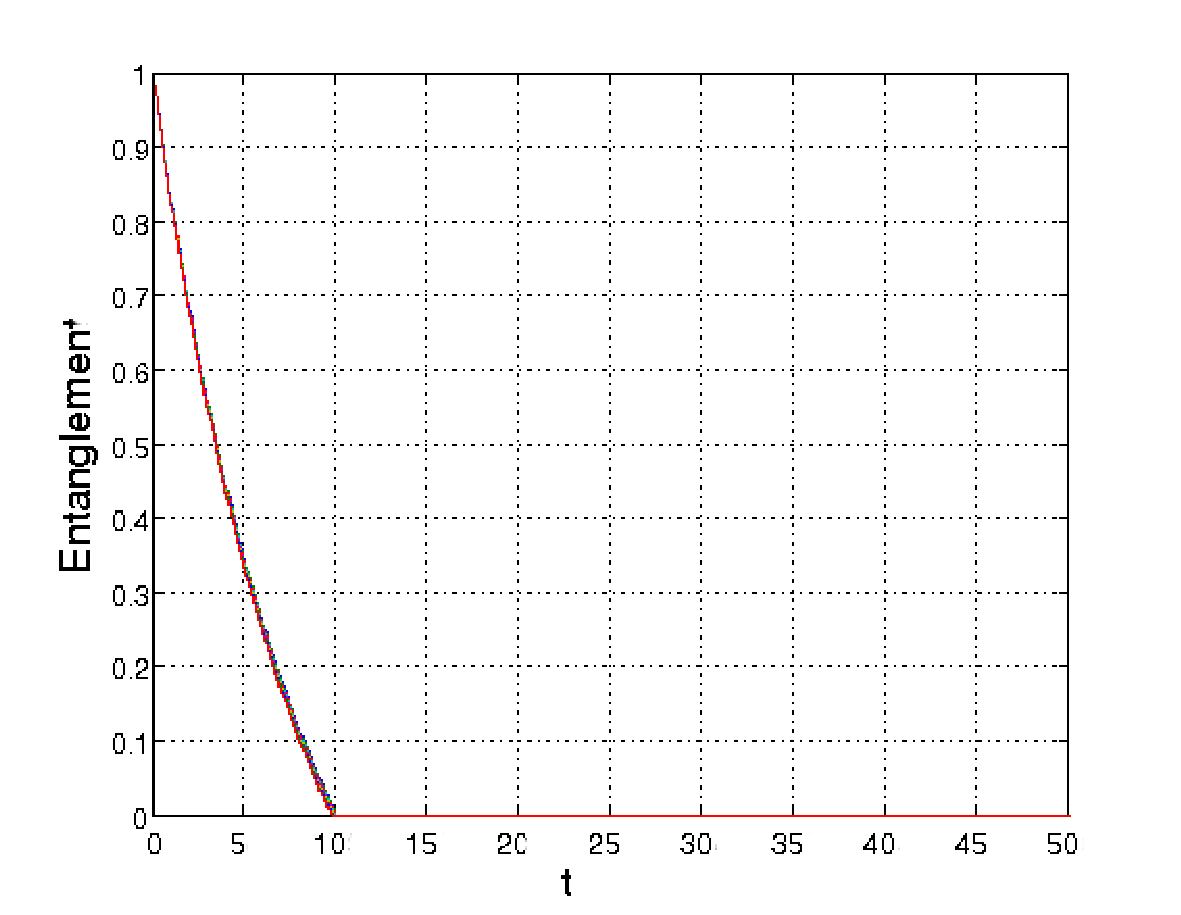}
 \caption{(Color online) Effect of bang-bang decoupling on  entanglement in the
   strong coupling  region. Here again we can see the effect of
 bang-bang decoupling on the entanglement if we compare this plot with
 FIG.(\ref{kappa_time_strong}). Here the parameters are same as in
 FIG. (\ref{kappa_time_strong}) and the pulse strength is $U =
 50\pi$ with the pulse duration $\tau = 0.01$.}\label{bang-strong-kappa} 
 \end{figure}

The Josephson charge qubit in contact with a $1/f$ bath is now subject to fast pulses, under the Bang-Bang dynamical decoupling scheme. The Hamiltonian for this radio frequency pulse is the same as in Eq. (\ref{pulse-hamil}). If the time for which a pulse is active is $\pi$, then the evolution operator for the pulse may be written as $V_p =  \mathbb{I}\otimes i\sigma_x$ where  $2U\tau_p = \pm \pi$. The total evolution
can therefore be written as
\begin{align}
V_{total} &= (V_pV_S(\tau))^{2N}
\end{align}
where $2N\tau = t$.  Since the RF pulses act on the system for very short amounts of time, the evolution of the system can safely be assumed to be governed only by the dynamical map $V_p$ for the time period during which the pulse is operating. As can be seen from the FIGS. \ref{bang-weak-kappa} to \ref{bang-strong-kappa} and FIG. \ref{cross-over} the system exhibits the  ESD on the application of bang-bang pulses.

Let us consider the case where $E_j = \epsilon = 1$. Let us also fix the pulse strength to be $U=50\pi$ and ensure that the pulses act for very short times. As defined earlier, the ratio of the BC bias $v$ and the switching rate $\gamma$ defines the weak and strong coupling regimes, the former designated by $\displaystyle \frac{v}{\gamma} \ll 1$ and the latter by $\displaystyle \frac{v}{\gamma} > 1$. In 
FIG.  \ref{bang-weak-kappa}, we plot the time-evolution of entanglement, with the coupling strength as parameter. For weak coupling, we find that $t_{ESD}$ initially increases with coupling strength. This continues till a turning point is reached at $\displaystyle \frac{v}{\gamma} = 0.38$ when $t_{ESD}\simeq 880$. After this, with increase in coupling strength, $t_{ESD}$ starts to decrease. 
As a result, a kink appears in the corresponding entanglement vs time plot, see Figs.~\ref{bang-weak-kappa},  \ref{bang-kink}.
The receding  of $t_{ESD}$ with increase in coupling strength continues well into the strong coupling regime, i.e. for $\displaystyle 5.05 < \frac{v}{\gamma} < 5.5$. 
It, however, does not go to zero, but rather chooses to saturate at the threshold value of $t_{ESD}\simeq 10$, see FIG. \ref{bang-strong-kappa}. The ``turning'' and the ``saturation'' 
features are well captured in FIG.  \ref{cross-over}, where we plot $t_{ESD}$ against $\displaystyle \frac{v}{\gamma}$ and keep the pulse strength and durations fixed. 
The saturation behavior is consistent with what one expects of $1/f$ noise. We observe a crossover phenomenon around  $\displaystyle \frac{v}{\gamma} \simeq 0.38$, where the value of $t_{ESD}$ rises sharply, only to fall back again even quicker. The crossover phenomena is a signature of the transient behavior exhibited by the system in going from the weak to the strong coupling regime. As discussed in the beginning of this subsection, saturation and transient behavior are characteristic of $1/f$ noise.

A SQUID (superconducting quantum interference device) is obtained by quantizing what is mathematically equivalent to a forced damped pendulum, and hence
a forced damped oscillator for a small superconducting phase \cite{ch}. Under the conditions $k_B T \ll E_j$ and $E_C \gg E_j$, and neglecting the damping as well as the biasing terms,
the Hamiltonian of Eq. (\ref{Josephson}) is obtained. Including the biasing provided by the BCs as well as the dissipation due to the bath, this problem could be thought of being
analogous to that of resonance flourescence, where the dissipative system is biased by an external field. In the present case, the biasing is embodied by the parameter $v$, due to the BC, 
and through it
by $\kappa$, while for resonance flourescence the biasing is from the external field and is quantified by the Rabi frequency $\Omega$. Due to the action of the Bang-Bang pulses, the dissipative
effect of the system ($\gamma$) is reduced and this suggests an  analogy between the Josephson junction charge qubit under the action of Bang-Bang pulses and the underdamped regime of
resonance flourescence. Indeed, with increase in value of $\kappa$, increase in ESD takes place and finally saturates for the case of strong coupling (large $\kappa$), in consonance
with a similar pattern in the case of resonance flourescence in the underdamped regime, where ESD is seen to increase with increase in $\Omega$ till a saturation is achieved. 

\begin{figure}
 \includegraphics[width=7.5cm]{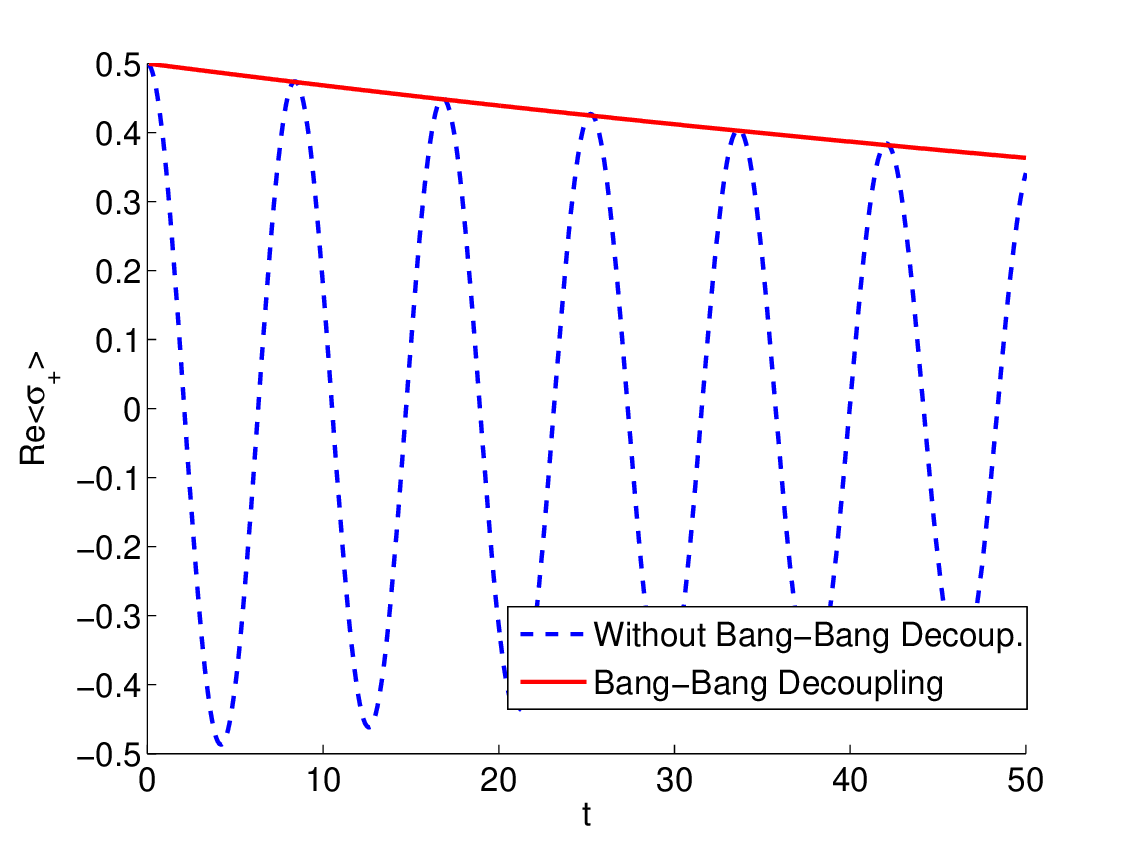}
\caption{(Color online) Plot for the evolution of coherence in the case of Telegraph noise in weak coupling region, i.e, $\kappa <1$. Here all the parameters has the value same as in FIGS. \ref{bang-kink} and \ref{bang-weak-kappa} and $\kappa = 0.38$.}\label{coh-tele-weak}
\end{figure}

\begin{figure}
 \includegraphics[width=7.5cm]{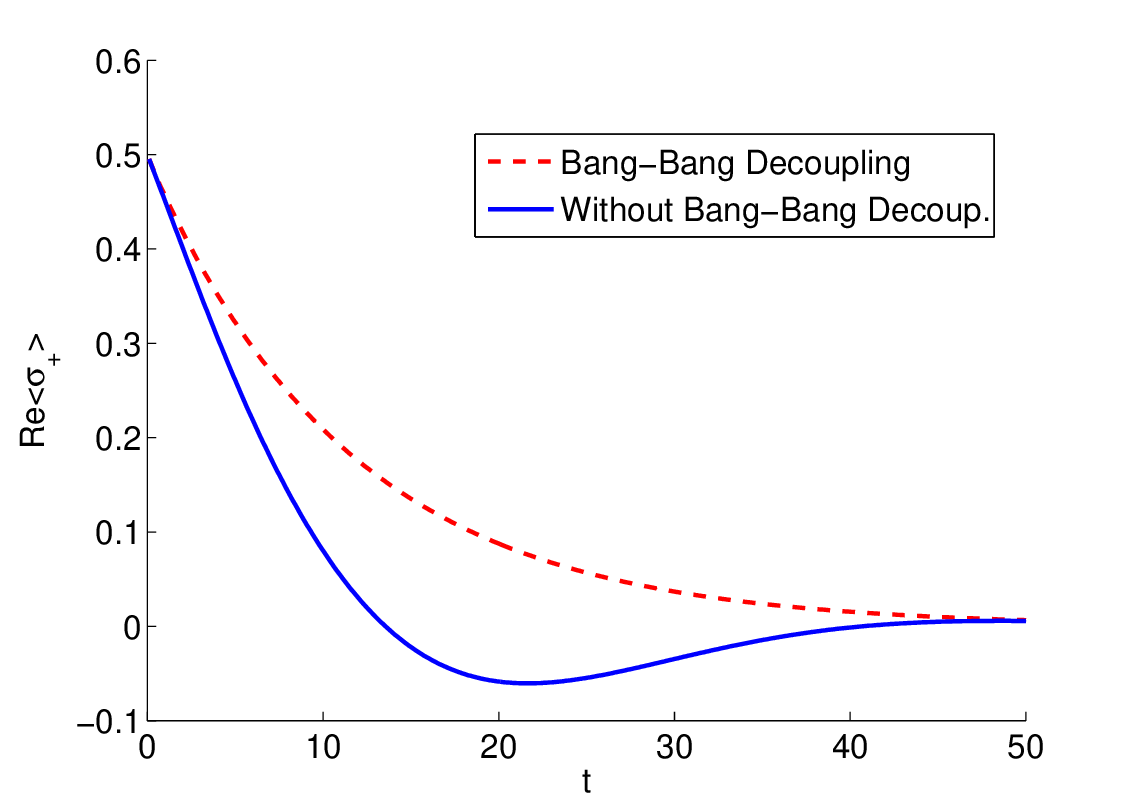}
\caption{(Color online) Plot for the evolution of coherence in the case of Telegraph noise in strong coupling region, i.e, $\kappa \ge 1$. Here the value of all the parameters are same as in Fig. \ref{bang-strong-kappa} and $\kappa = 5.38$.}\label{coh-tele-strong}
\end{figure}

\begin{figure}
 \includegraphics[width=7.5cm]{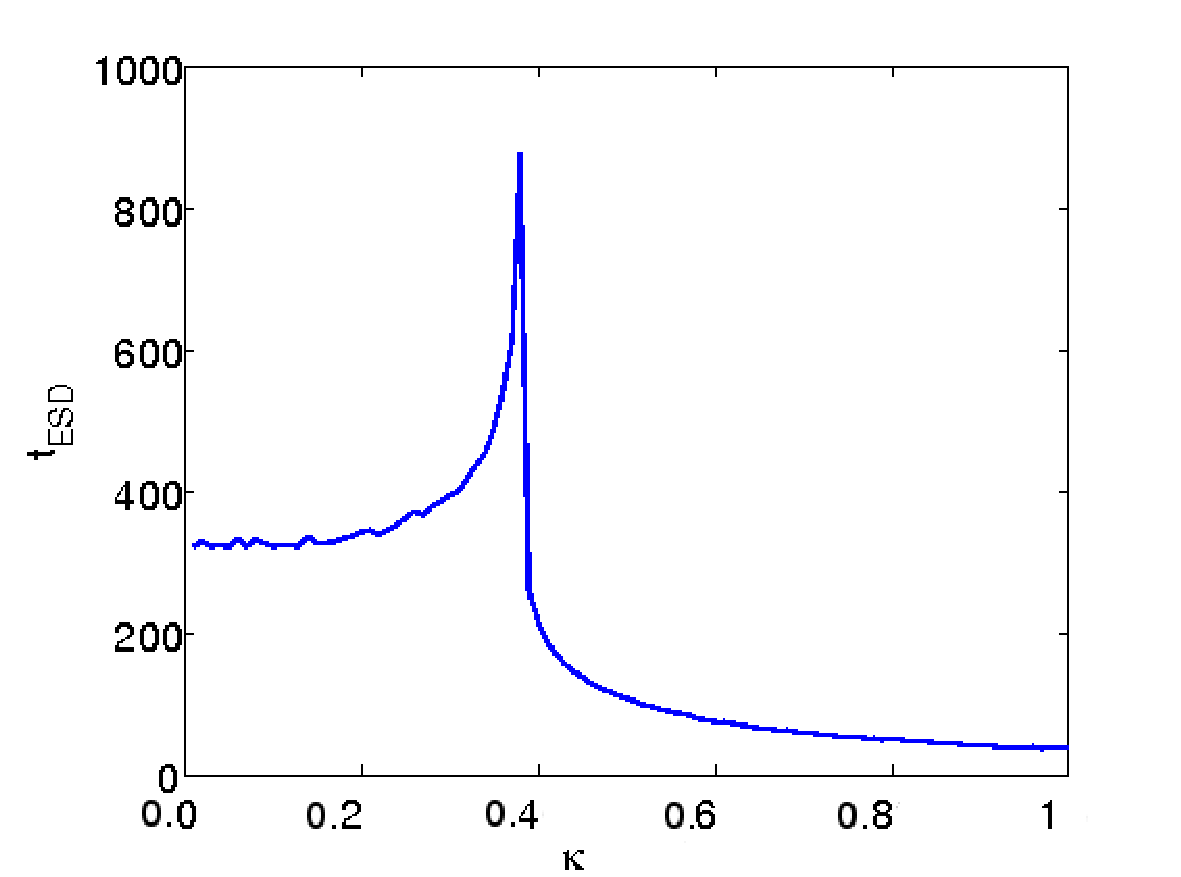}
 \caption{(Color online) $t_{ESD}$ is plotted as a function of coupling strength
   $\kappa$. Here we can see that there is a clear distinction between the
   strong  and the weak coupling region. As we increase
 $\kappa$ the $t_{ESD}$ tends to freeze and asymptotic value of $t_{ESD}$ is
 around $10$. The parameters used are as in the previous plots.}\label{cross-over}
 \end{figure}

The evolution of coherence with respect to time, for the Josephson charge qubit subjected to $1/f$ noise,  is shown in FIGS. \ref{coh-tele-weak},   \ref{coh-tele-strong},  for the weak and strong coupling regimes, respectively.  Both show an improvement in the coherence with the application of the bang-bang decoupling pulses, in contrast to the corresponding behavior of entanglement, thereby reiterating that coherence is not synonymous with entanglement.

\begin{figure}
 \includegraphics[width=7.5cm]{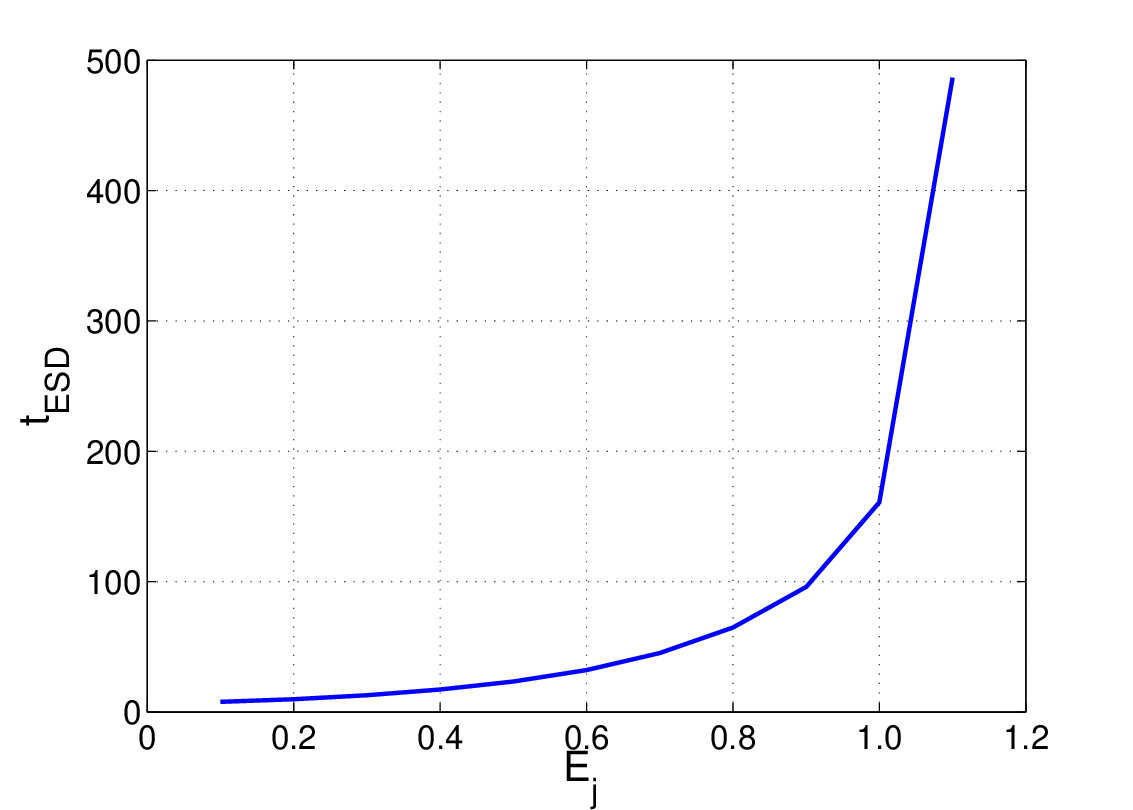}
 \caption{(Color online) $t_{ESD}$ is plotted as a function of $E_j$. This shows that
   as we go away from  pure dephasing (i.e, $E_j=0$) the $t_{ESD}$ increases. The parameters used are same as in the previous plots.}\label{ent-ej}
 \end{figure}

In FIG.  \ref{ent-ej}, we plot the behavior of $t_{ESD}$ with $E_j$ and find that, as we increase $E_j$ and thus move away from the pure dephasing situation, the time to ESD keeps increasing. 
As discussed earlier, this is a counterintuitive result because dissipation increases with $E_j$. This may be explained by invoking the results of FIG. 7: as $E_j$ increases, $\epsilon$ kept fixed, entanglement increases, which in turn implies increase in time to ESD.


\section{Conclusions}

The importance of the sustenance of entanglement in quantum systems cannot be overstated. In this paper, we have studied a variety of control procedures aimed at doing exactly that. 
Most of these are designed to suppress decoherence at the level of single qubits. 
A majority of the systems considered in this paper are qubits coupled with harmonic oscillator baths at finite temperature $T$, the couplings being either of dissipative or of the dephasing type. 
The time-evolution of entanglement when such a bath acts on one side of the two-qubit maximally entangled state is known \cite{SKG10}. In the commonly occuring dissipative case, entanglement decays asymptotically at zero temperature, whereas it shows a sudden death at finite non-zero temperatures. Squeezing in the initial bath states increases the time to ESD.

The aim of most control procedures is to suppress decoherence. For the case of photonic crystals, the design allows the system to conserve coherence when it is within the photonic band gap. Modulating the frequency of the system-bath coupling aims to suppress decoherence by shifting the system out of the spectral influence of the bath. In both these cases it is found that the suppression of decoherence is accompanied by a corresponding increase in $t_{ESD}$. 

However, it will be erroneous to na\"{i}vely suppose that this is the norm. Exactly the  opposite phenomenon is observed for the case of resonance fluorescence, where the coupling between the 
bath and a two-level atomic system forced by an external resonant field, is modulated. It is seen that an increase in the external field frequency $\Omega$, the Rabi frequency, results in a 
faster decay of entanglement (FIG. \ref{concurrenceVstime}). A further non-trivial effect observed is the saturation in the time to ESD: $t_{ESD}$ does not go below a threshold value no matter 
what the Rabi frequency. A possible explanation of this phenomenon would be to observe that the sudden death time stops being dependent on the Rabi frequency 
at $\displaystyle \Omega = \frac{\gamma_0}{4}$; strikingly, this happens to be the boundary between the overdamped $\displaystyle \Omega < \frac{\gamma_0}{4}$ and 
underdamped $\displaystyle \Omega > \frac{\gamma_0}{4}$ regimes.

\begin{widetext}
\begin{center}
\begin{table}
\begin{tabular}{|c|c|c|}
\hline
Control Procedure & Decoherence suppression & Entanglement decay suppression \\
\hline
\hline 
Photonic Crystals & \checkmark & \checkmark\\
Frequency Modulation &  \checkmark & \checkmark\\
Resonance Fluorescence & \checkmark & $\times$\\
DD in EMF bath &  \checkmark & \checkmark\\
DD in Telegraph noise &  \checkmark & $\times$\\
\hline
\end{tabular}
\caption{Summary of the results.}\label{summery}
\end{table}
\end{center}
\end{widetext}

In dynamical decoupling schemes  RF pulses, applied at short time-intervals, smooth out unwanted effects due to environmental interactions. We discuss two qualitatively different system-bath 
models: the first being the usual qubit and harmonic oscillator bath pair with pure dephasing or QND interaction; and the second being a bath of charge impurities, simulating $1/f$ (telegraph) 
noise, acting on a Josephson-junction charge qubit. Entanglement decays to zero asymptotically in both these models. The application of fast RF pulses to the former manages to speed up the rate 
of the still-asymptotic loss of entanglement, whereas the same RF pulses applied to the latter kills off entanglement in finite time and thus shows ESD. A very interesting phenomenon, observed in 
the strong coupling regime, is the decrease in the time to ESD with increasing pulse strengths. This is extremely counterintuitive, and brings into perspective the fact that, in the
 non-Markovian strong coupling regime, the dynamics of entanglement can be different than that of decoherence. This feature gets further highlighted by the behavior of coherence with time, 
both for the case of resonance fluorescence and Josephson-junction charge qubit subjected to $1/f$ noise. Here coherence -- which is  a local property -- is seen to vary in a  non-monotonic 
fashion with entanglement which happens to be  a non-local property of the system. A summary of our results is presented in tabular form in Table I.
This, thus, calls for the need to have careful and exhaustive studies of entanglement in these regimes.

\acknowledgments

We wish to thank Somdeb Ghose for his suggestions to improve the readability of the article.  SB thanks T. P. Pareek for a useful discussion.

\appendix
\section{Calculation for concurrence}
\label{concurrence}
Concurrence is a measure for entanglement of formation for a mixed
state of two-qubit system given by Hill {\em et al.}
\cite{Hill97,Wootters98}. Concurrence is defined as $\mathcal{C} = max(0,
\lambda_1-\lambda_2-\lambda_3-\lambda_4)$, where $\{\lambda_i\}$ are
the square root of the eigenvalues of the matrix $R =
\rho\tilde{\rho}$. Here $\tilde{\rho} = \left(\sigma_2\otimes \sigma_2\right)
\rho^* \left(\sigma_2\otimes \sigma_2\right)$ and the complex conjugate is taken in
the standard basis.

For some simple density matrices we can calculate the concurrence very
easily. For example, consider the matrix
\begin{align}
M &= \frac{1}{2}\left(\begin{array}{cccc}
1& 0& 0& e^{-\gamma t}\\
0 &0&0&0\\
0 &0&0&0\\
e^{-\gamma t}&0&0&1
\end{array}\right).
\end{align}
The $R$ matrix will be:
\begin{align}
R &= M\tilde{M}
\end{align}
and 
\begin{align}
\tilde{M} &= \sigma_2\otimes \sigma_2
M^* \sigma_2\otimes \sigma_2\\
&= \frac{1}{2}\left(\begin{array}{cccc}
1& 0& 0& e^{-\gamma t}\\
0 &0&0&0\\
0 &0&0&0\\
e^{-\gamma t}&0&0&1
\end{array}\right).\\
R &= \frac{1}{4}\left(\begin{array}{cccc}
1+e^{-2\gamma t} & 0& 0& 2e^{-\gamma t}\\
0 &0&0&0\\
0 &0&0&0\\
2e^{-\gamma t}&0&0&1 + e^{-2\gamma t}
\end{array}\right).
\end{align}
The set $\{\lambda_i\}$  is equal to $\{ (1+e^{-\gamma t})/2,~ (1-e^{-\gamma
  t})/2\}$. The concurrence for this state is $\mathcal{C} = e^{-\gamma t}$.

\section{ Calculation of $V_{fm}$ and $M_{fm}$ for Frequency modulation}
\label{app-fm}
In the case of Frequency modulation, from Eq.~\eqref{masterdriven} we can write the $L_{fm}$ and thus the $V_{fm}= \exp(L_{fm}t)$ matrix as:
\begin{widetext}
\begin{align}
L_{fm} &= \left(\begin{array}{cc|cc}
-2{\rm Re}(\alpha) C_0^{-+} &0 &0 &2{\rm Re}(\alpha) C_0^{+-}\\
0&-\alpha(C_0^{-+}+C_0^{+-})&0&0\\
\hline
0&0&-\alpha^*(C_0^{-+}+C_0^{+-})&0\\
2{\rm Re}(\alpha) C_0^{-+}&0 &0 & -2{\rm Re}(\alpha) C_0^{+-}
\end{array}\right),\\
V_{fm} & = \left(\begin{array}{cccc}
\frac{1}{T}\left(C_0^{-+}e^{-2{\rm
  Re}(\alpha)Tt}+C_0^{+-}\right)&0&0&\frac{1}{T}\left(C_0^{+-}(1-e^{-2{\rm
  Re}(\alpha)Tt})\right)\\
0&e^{-\alpha Tt}&0&0\\
0&0&e^{-\alpha^*Tt}&0\\
\frac{1}{T}\left(C_0^{-+}(1-e^{-2{\rm
  Re}(\alpha)Tt})\right)&0&0&\frac{1}{T}\left(C_0^{+-}e^{-2{\rm
  Re}(\alpha)Tt}+C_0^{-+}\right)\end{array}\right),
\end{align}
\end{widetext}
where $\alpha =
\frac{2(\kappa-i\Delta)J_1^2(m)}{(\kappa-i\Delta)^2+\nu^2}$ and $T = C_0^{-+} + C_0^{+-}$.
If $M_{fm} = (\mathbb{I}\otimes V)(|\phi^+\rangle\langle \phi^+|)$, then we have
\begin{align}
M_{fm} = \left(\begin{array}{cccc}
M_{11}&0&0&e^{-\alpha Tt}\\
0&M_{22}&0&0\\
0&0&M_{33}&0\\
e^{-\alpha^*Tt}&0&0&M_{44}\end{array}\right),
\end{align}

where
\begin{align}
 M_{11} &= \frac{1}{T}\left(C_0^{-+}e^{-2{\rm  Re}(\alpha)Tt}+C_0^{+-}\right),\\
M_{22} &= \frac{1}{T}\left(C_0^{+-}(1-e^{-2{\rm  Re}(\alpha)Tt})\right),\\
M_{33} &= \frac{1}{T}\left(C_0^{-+}(1-e^{-2{\rm  Re}(\alpha)Tt})\right),\\
M_{44} &= \frac{1}{T}\left(C_0^{+-}e^{-2{\rm  Re}(\alpha)Tt}+C_0^{-+}\right).
\end{align}
If $M_{fm}$ is separable at some time $t$, the factorization law for entanglement decay \cite{Konrad08} allows us to assert that all states will show
ESD. The state $M_{fm}$  is separable if only if  it is positive under partial transposition, i.e,
\begin{align}
1 + X^2 -2X -\frac{T^2}{C_0^{-+}C_0^{-+}}X \ge 0,\label{B8}
\end{align}
, where $X= \exp(-2{\rm Re}(\alpha)Tt)$. Therefore, $M_{fm}$ is separable when LHS of Eq. \ref{B8} is zero. The roots of the above equation are
\begin{align}
X_{\pm} = \frac{1}{2} \left[ \left( 2+\frac{T^2}{C_0^{-+}C_0^{-+}} \right) \pm \sqrt{\left(2+\frac{T^2}{C_0^{-+}C_0^{-+}}\right)^2-4} \right].
\end{align}
The negative root is less than unity ($X_- \le 1$), implying that there exists, always, a finite and positive time $t_{ESD}$ at which the system loses all its entanglement. This is given by
\begin{align}
t_{ESD} = -\frac{1}{2{\rm Re}(\alpha)T}\log (X_-)\label{B10}.
\end{align}
The modulation factor $\nu$ appears in the numerator of Eq. \ref{B10} and, therefore, it can be expected that a higher frequency of modulation should sustain entanglement longer. This is confirmed in the plot of $t_{ESD}$ against $\nu$ (FIG. \ref{frequency_modulation}). This result is not altogether surprising, for a higher degree of modulation is naturally expected to increase the coherence  by filtering out the influence of the bath, which ultimately results in entanglement sustaining for a longer period of time.

\section{ Calculation for the evolution map $V_s$ acting on the
  qubit from the evolution map of the qubit plus charge-impurity in the case of telegraph noise}
\label{app-tele}
The matrix representation of the evolution map for qubit plus
charge-impurity $V$ is given by $\exp(Rt)$ where $R$ is given in
Eq. (\ref{redfield-tensor}). This map is in the basis $\{
|\theta_{\pm}\rangle|i\rangle\}\otimes\{
|\theta_{\pm}\rangle|j\rangle\}$. To make it computationally easier we
need to write it in the basis
$\{|\theta_{\pm}\rangle|\theta_{\pm}\rangle\}\otimes
\{|i\rangle|j\rangle\}$, since the matrix representation of the map
acting on the qubit is in the basis
$\{|\theta_{\pm}\rangle|\theta_{\pm}\rangle\}$. To change the basis we
need the assistance of a unitary matrix (in this case permutation
matrix) $P$ which is defined as:

\begin{align}
P(|a\rangle|b\rangle|c\rangle|d\rangle) &=
|a\rangle|c\rangle|b\rangle|d\rangle; \\
|a\rangle|b\rangle|c\rangle|d\rangle &=
\left( \begin{array}{c}a_1\\a_2\end{array}\right)
\otimes
\left( \begin{array}{c}b_1\\b_2\end{array}\right)
\otimes
\left( \begin{array}{c}c_1\\c_2\end{array}\right)
\otimes 
\left( \begin{array}{c}d_1\\d_2\end{array}\right),\\
|a\rangle|c\rangle|b\rangle|d\rangle &= 
\left( \begin{array}{c}a_1\\a_2\end{array}\right)
\otimes
\left( \begin{array}{c}c_1\\c_2\end{array}\right)
\otimes
\left( \begin{array}{c}b_1\\b_2\end{array}\right)
\otimes 
\left( \begin{array}{c}d_1\\d_2\end{array}\right).\\
\Rightarrow P & = \mathbb{I}\otimes p \otimes \mathbb{I},
\end{align}
where
\begin{align*}
p &= \left(\begin{array}{cccc}
1 & 0 & 0 & 0\\
0 & 0 & 1 & 0\\
0 & 1 & 0 & 0\\
0 & 0 & 0 & 1
\end{array}\right).
\end{align*}
After conjugating the matrix $V$ by $P$ we get:
\begin{align}
\tilde{V} &= PVP^T.
\end{align}
We can write this $16\times 16$ matrix $\tilde{V}$ as a $4\times 4$
matrix, where each of the element itself is a $4\times 4$ matrix
$Z_{ij}$ where $i,j \in \{1,2,3,4\}$. Then the map $V_s$ acting on
qubit is simply $V_{s\, ij} = {\rm Tr}(Z_{ij})$. From here we can get
the corresponding $M$ matrix.
It is not easy to solve it analytically in the present case. Therefore, we use  
numerical methods to calculate the
evolution operator and  entanglement evolution for a system of
qubits.


\end{document}